\documentclass[transmag]{IEEEtran}
\usepackage{latexsym}
\usepackage{graphicx}
\usepackage{amsfonts,amssymb,amsmath}
\usepackage{hyperref}
\usepackage[export]{adjustbox}
\usepackage{array}
\usepackage{float}
\usepackage{makecell}
\usepackage{placeins}
\usepackage{graphicx}
\DeclareGraphicsExtensions{.pdf,.jpeg,.png,.eps}
\usepackage{algorithm}
\usepackage{textcomp}
\usepackage{algpseudocode}
\usepackage{supertabular}
\usepackage{listings}
\usepackage{multirow}
\usepackage{amsmath}
\usepackage{mathtools}
\usepackage{epstopdf}
\usepackage{authblk}
\usepackage{subfigure}
\usepackage{cite}
\usepackage{cleveref}
\usepackage{latexsym}
\usepackage[utf8]{inputenc}
\usepackage{amsthm}
\usepackage{balance}
\usepackage{relsize}
\usepackage[nodisplayskipstretch]{setspace}
\usepackage{amssymb}
\usepackage{eucal}
\usepackage{url}
\usepackage{color}  
\definecolor{Mypink}{RGB}{255,0,255}
\definecolor{Myorange}{RGB}{255,102,0}
\definecolor{Mygreen}{RGB}{0,153,0}
\definecolor{Myblue}{RGB}{0,0,255}
\usepackage{lipsum}

\DeclareMathAlphabet\mathbfcal{OMS}{cmsy}{b}{n}
\DeclareUnicodeCharacter{2061}{}

\def\BibTeX{{\rm B\kern-.05em{\sc i\kern-.025em b}\kern-.08em T\kern-.1667em\lower.7ex\hbox{E}\kern-.125emX}}

\begin{document}

\title{Rate-Splitting Multiple Access: Unifying NOMA and SDMA in  MISO VLC Channels \\ \textit{(Invited Paper)}}

\renewcommand\Authfont{\fontsize{12}{14.4}\selectfont}
%\renewcommand\Affilfont{\fontsize{12}{14.4}\selectefont}

%\author{Shimaa~Naser, Lina~Bariah, Wael~Jaafar, Sami~Muhaidat, Paschalis~C.~Sofotasios, Mahmoud~Al-Qutayri, and Octavia~A.~Dobre
\author{Shimaa Naser, Lina Bariah, \IEEEmembership{Member, IEEE}, Wael Jaafar, \IEEEmembership{Senior Member, IEEE}, \\Sami Muhaidat, \IEEEmembership{Senior Member, IEEE}, Paschalis C. Sofotasios, \IEEEmembership{Senior Member, IEEE}, \\Mahmoud Al-Qutayri, \IEEEmembership{Senior Member, IEEE}, and Octavia A. Dobre, ~\IEEEmembership{Fellow,~IEEE}

\thanks{S. Naser, L. Bariah and S. Muhaidat are with the Center for Cyber-Physical Systems, Department of Electrical Engineering and Computer Science, Khalifa University, Abu Dhabi, UAE, (e-mails: 100049402@ku.ac.ae; lina.bariah@ku.ac.ae;muhaidat@ieee.org.)}
\thanks{P. C. Sofotasios is with the Center for Cyber-Physical Systems, Department of Electrical Engineering and Computer Science, Khalifa University, Abu Dhabi, UAE, and also with the Department of Electrical Engineering, Tampere University, Tampere, Finland, (e-mail: p.sofotasios@ieee.org.)}
\thanks{M. Al-Qutayri is with the Department of Electrical Engineering and Computer Science, Khalifa University, Abu Dhabi, UAE, (e-mail: mahmoud.alqutayri@ku.ac.ae
.)}
\thanks{W. Jaafar is with the Department of Systems and Computer Engineering, Carleton University, Ottawa, ON, Canada, (e-mail: waeljaafar@sce.carleton.ca.)}
\thanks{O. A. Dobre is with the Faculty of Engineering and Applied Science University, Memorial University, St. John's, Canada, (e-mail: odobre@mun.ca.)}
}

\IEEEtitleabstractindextext{\begin{abstract}
The increased proliferation of connected devices requires a paradigm shift towards the development of innovative technologies for the next generation of wireless systems. One of the key challenges, however, is the spectrum scarcity, owing to the unprecedented broadband penetration rate in recent years. Visible light communications (VLC) has recently emerged as a possible solution to enable high-speed short-range communications. However, VLC systems suffer from several limitations, including the limited modulation bandwidth of light-emitting diodes. Consequently, several multiple access techniques (MA), e.g., space-division multiple access (SDMA) and non-orthogonal multiple access (NOMA), have been considered for VLC networks. Despite their promising multiplexing gains, their performance is somewhat limited. In this article, we first provide an overview of the key MA technologies used in VLC systems. Then, we introduce rate-splitting multiple access (RSMA), which was initially proposed for RF systems and discuss its potentials in VLC systems. Through system modeling and simulations of an RSMA-based two-use scenario, we illustrate the flexibility of RSMA in generalizing NOMA and SDMA, as well as its superiority in terms of weighted sum rate (WSR) in VLC. Finally, we discuss challenges, open issues, and research directions, which will enable the practical realization of RSMA in VLC.

\end{abstract}


}
\maketitle

\section{INTRODUCTION}
 \IEEEPARstart{T}{he} exponential growth of connected devices and  emergence of the Internet-of-Everything (IoE), enabling ubiquitous connectivity among billions of people and machines, have been the major driving forces towards the evolution of wireless technologies, aiming  to support a plethora of new services, including  enhanced mobile broadband and ultra-reliable and low-latency communications. While the demand for new IoE services, e.g., extended reality, autonomous driving and tactile Internet continues to grow, it is necessary for future wireless networks to deliver high reliability, low latency, and very high data rates. In this context, the notion of visible light communications (VLC) has emerged as a promising wireless technology for massive connectivity of users with high data rates. %and massive connectivity to users through fast fiber backhauling. \\
  
\begin{figure*}[ht]
%\begin{figure}
\centering
\includegraphics[width=1\linewidth]{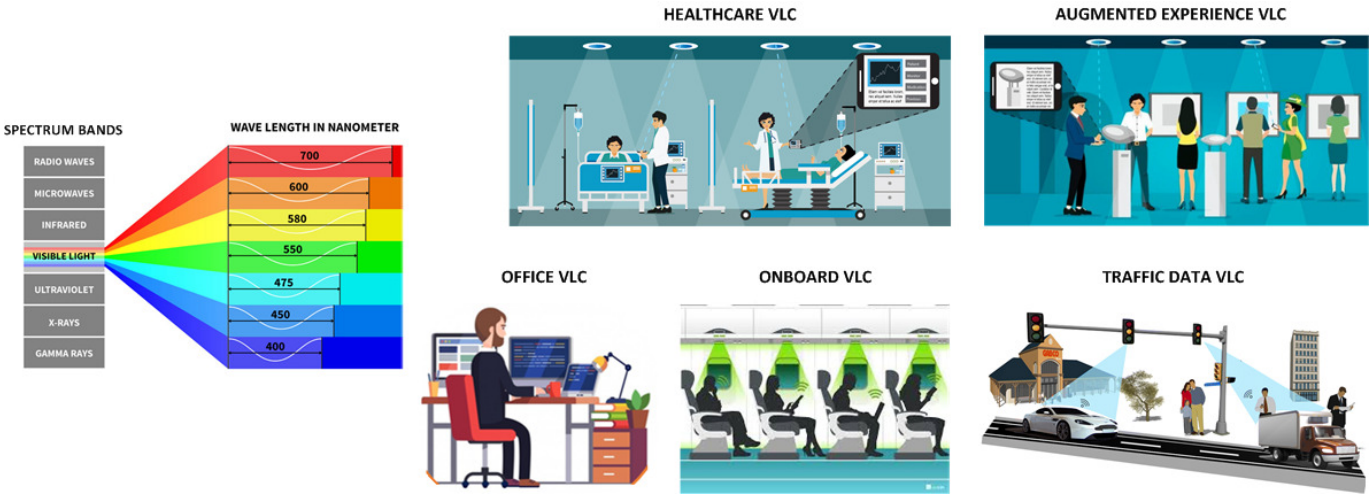}
%\centerline{\includegraphics[width=3.5in]{OJVT/Fig1.png}}

\caption{VLC spectrum and examples of its use.}
\label{Fig:VLC1}
\end{figure*}

To realize VLC, a  simple  and  inexpensive  modification is required to the existing lighting infrastructure \cite{Komine2004,Elgala2011,Ghassem2019,Ndjiongue}. The key attractive features of VLC include, but are not limited to, security, high degree of spatial reuse, and immunity to electromagnetic interference \cite{Grobe2013}. The advancement in solid-state has introduced light emitting diodes (LEDs) as  energy-efficient light sources, which are envisioned  to dominate the next generation of wireless infrastructure. One of the interesting features of LEDs is their ability to rapidly switch between different light intensities in a way that is not perceptible to human eyes. This enables them to be the main technology for VLC systems. The key principle of VLC is to use  emitted light from the LEDs to perform data transmission through intensity modulation and direct detection (IM/DD), without affecting the LEDs' main illumination function. 
%It would be emphasized that, the utilization of LEDs for data transfer should not have any negative impact on their illumination function. 
The huge unregulated spectrum of visible light allows VLC to offload data traffic from radio-frequency (RF)/microwave systems while providing high data rates.  VLC uses the 400 THz to 789 THz visible light spectrum, which is characterized by the features of low penetration  through objects, secure communications, and high quality-of-service (QoS) in interference-free small cells designs \cite{Karuna2015,Mirami2015,IEEE2018}. Fig. \ref{Fig:VLC1} shows the VLC spectrum band and examples of its use in healthcare, work office, transportation, and smart cities. Also, the structure of the survey is illustrated

\begin{table*}[t]
\scriptsize
\centering
\caption{List of acronyms and Abbreviations.}
\begin{tabular}[l] {|p{2.5cm}|p{6cm}|p{2.5cm}|p{5cm}|}
\hline
\textbf{Abbreviation }&\textbf{Definition}&\textbf{Abbreviation}&\textbf{Definition }  \\ \hline \hline 
%ACO-OFDM& 	Asymmetrically Clipped Optical OFDM\\ \hline
ADT & Angle Diversity Transmitter& MU 	&Multi-User \\ \hline

%ADO-OFDM&	Asymmetrically Clipped DC Biased Optical OFDM 
AO	&Alternating Optimization &{MUD} &Multi-User Detection \\ \hline  

BD &	Block Diagonalization & MUI	&Multi-User Interference\\ \hline 

BER &	Bit Error Rate & NOMA 	&Non-Orthogonal Multiple Access \\ \hline

%&OOC& optical orthogonal codes
BC&	Broadcast Channel & OCDMA &Optical Code-Division Multiple Access	\\ \hline

CDMA &	Code-Division Multiple Access &OFDM	&  Orthogonal  Frequency-Division Multiplexing\\ \hline

C-NOMA&	Code NOMA &OFDMA&Orthogonal Frequency-Division Multiple Access \\ \hline 

CoMP 	&Coordinated multi-point &OMA	& Orthogonal Multiple Access\\ \hline

CSI 	&Channel State Information & OOK	&On-Off Keying \\ \hline

CSIT 	&CSI at Transmitter &  P-NOMA	&Power NOMA\\ \hline

%PAM & Pulse Amplitude Modulation  \\ \hline 

%CSK 	&Color Shift Keying 

DC 	& Direct Current &PD&Photo Detectors \\ \hline
%DCO-OFDM& 	DC Biased Optical OFDM & 

DD 	&Direct Detection &QoS& Quality-of-Service\\ \hline

FoV 	& Field of View &RGB	& Red, Green and Blue \\ \hline 

ICI & Inter-Channel Interference &RSMA	&Rate-splitting Multiple Access\\ \hline 
IM 	& Intensity Modulation &SC	& Super-position Coding\\ \hline
ISI 	&Inter-Symbol Interference &SDMA &Space Division Multiple Access\\ \hline

LED 	&Light Emitting Diode& SIC & Successive Interference Cancellation \\ \hline

%SMP	&Spatial Multiplexing\\ \hline
LTE & Long-Term Evolution &SINR& Signal-to-Interference-Plus-Noise Ratio\\  \hline

LoS &Line-of-Sight &SNR &Signal-to-Noise Ratio\\ \hline

MA	& Multiple Access &TDMA& Time-Division Multiple Access \\ \hline

MAC & Media Access Control  &VLC&	Visible Light Communication \\ \hline

MIMO &	Multiple-Input Multiple-Output& WSMSE	&Weighted Sum Mean Squared Error \\  \hline

MISO	&Multiple-Input Single-Output &WSR&Weighted Sum Rate\\  \hline

MMSE 	& Minimum Mean-Square Error &ZF	& Zero-Forcing \\ \hline

MSE 	&Mean-Square Error & ZF-DPC 	&Zero-Forcing Dirty-Paper\\ \hline

%WMMSE	&Weighted Minimum Mean Squared Error\\
%WSMSE	&Weighted Sum Mean Squared Error\\%\hline
%WSR	&Weighted Sum Rate \\ %\hline
%ZF & Zero-Forcing \\
%
\end{tabular}
\label{Table1}
\end{table*}

\begin{figure}[ht]
\centerline{\includegraphics[width=3.5in]{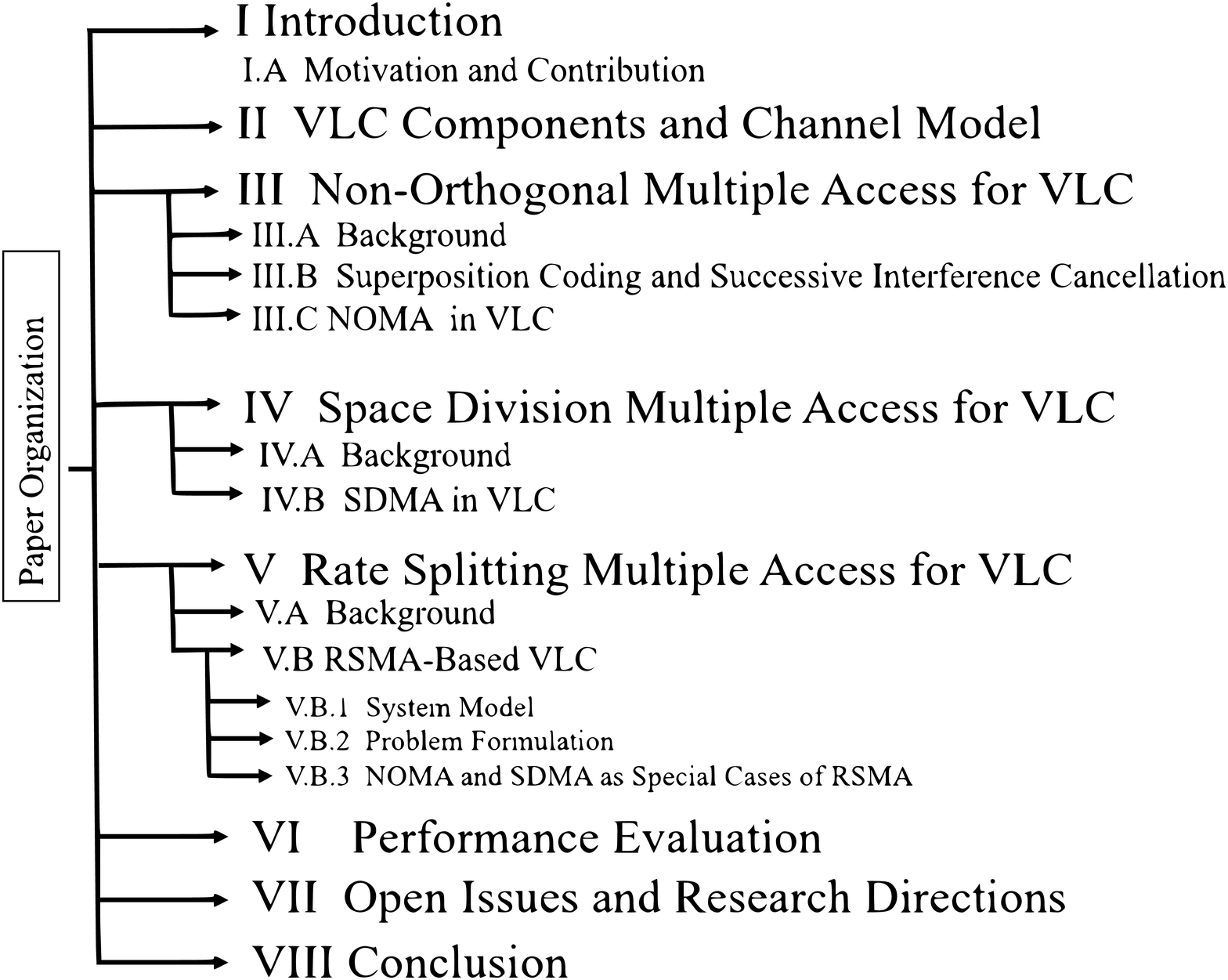}}
\caption{Organization of the paper.}
\label{Fig:chart}
\end{figure}

in Fig. \ref{Fig:chart}. 
\subsection{Motivation and Contribution}
Despite its advantages, VLC suffers from several drawbacks that limit its performance. For example, the limited modulation bandwidth and peak optical power of LEDs are considered as the main obstacles towards realizing the full potential of VLC systems \cite{Jovicic2013}. Therefore, several studies  have been carried out to enhance the spectral efficiency of VLC systems. In particular, two research directions have been identified: in the first one, researchers focused on the design of dedicated VLC analog hardware and digital signal processing techniques. In the second one, researchers focused on enhancing the spectral efficiency through the development of different optical-based modulation and coding schemes, adaptive modulation, equalization, VLC cooperative communications, orthogonal and non-orthogonal multiple access (OMA/NOMA) schemes, and multiple-input multiple-output (MIMO) \cite{Pathak2015}. 

In the context of VLC, several optical OMA schemes have been proposed, including time-division multiple access (TDMA), orthogonal frequency-division multiple access (OFDMA) \cite{Armstrong2009}, and optical code-division multiple access (OCDMA) \cite{Salehi1989}. These schemes rely on assigning orthogonal resources to different users. For example, in TDMA, different users are allocated different time slots for communication, while in OFDMA different users are assigned different orthogonal frequency sub-carriers. In  OCDMA, users communicate at the same time and frequency, which can be achieved through  the use of different orthogonal optical codes. In contrast, space-division multiple access (SDMA) exploits the spatial separation between users to provide full time and frequency resources. On the contrary, NOMA has been recently introduced as a spectrum-efficient multiple access (MA) scheme that allows different users to share the same time and frequency resources, leading to an enhanced spectral efficiency \cite{Ding2014}. NOMA is realized either by assigning different power levels to different users (known as power (P)-NOMA) or by allocating different spreading sequences (called code (C)-NOMA). Resource allocation in NOMA is determined according to different criteria, such as link quality, users fairness, targeted individual and sum rates, and users' QoS requirements.  

In the same context, rate-splitting multiple access (RSMA) has recently emerged as a potentially robust and generalized MA scheme for future wireless systems, which is able to accommodate different users in a heterogeneous environment. In particular, novel research results have shown that RSMA in MIMO-based RF systems outperforms other common MA schemes, such as NOMA and SDMA, in terms of spectral efficiency \cite{Mao2018}. \textcolor{black}{The performance gain of RSMA comes from the fact that the transmitted signal of each user is divided into one or several common parts and a private part. All common parts are multiplexed and encoded into a single (or several) common streams intended for all (or to a subset of) users. On the other hand, the private parts are encoded separately into  multiple private streams, which are then superimposed with the common stream(s). The super-symbol is then transmitted to all users over the VLC downlink. At each user, the common streams are decoded first in order to obtain the common parts of the intended user, utilizing iterative  successive interference cancellation (SIC). Subsequently, the private part is decoded while treating the other users' private parts as noise. The NOMA scheme can be obtained from RSMA by treating some users' signals as common parts and the remaining as private parts. On the other hand, SDMA can be realized from RSMA by using only the private parts to encode users' messages.} The split of the messages into common and private parts enables RSMA to provide robust services for different network loads and users deployments. 

In VLC systems, MIMO channels are practically highly correlated, which inevitably degrades the performance of linear precoding schemes. This has motivated the investigation of different receiver structures and precoding schemes in order to mitigate the effect of channel correlation in MIMO VLC systems. Likewise, several articles on the previously presented topics, namely OMA, NOMA and RSMA techniques design, have been presented for RF systems. Yet,  only few have addressed their application in VLC networks (mainly for OMA and NOMA), and summarized these studies in surveys \cite{Bawazir2018,Vaezi2019,Obeed2019}. Moreover, none of them has discussed the integration of RSMA into VLC systems. Motivated by the above, in this survey, we shed the light on several spectrally efficient MA schemes for VLC systems. In more details, we present a comprehensive study of NOMA and SDMA schemes, with particular attention to MIMO-VLC systems. In addition, we address the potential integration of the RSMA scheme in MIMO-VLC systems. Finally, open issues and some interesting related research directions are discussed.

\textit{Notation:} Bold upper-case letters denote matrices and bold lower-case letters denote vectors. $(\cdot )^{T}$ denotes the transpose operation, $\mathbb{E}(\cdot)$ is the statistical expectation operation, $|\cdot|$ is the absolute value operation, \textbf{\rm{I}} is the identity matrix, \textbf{0} is the zero matrix, tr$(\cdot)$ is the trace of a matrix, 
and $\mathcal{N}(0,\sigma^2)$ is a real-value Gaussian distribution with zero mean and variance $\sigma^{2}$. Let $\textbf{z}=[z_1,\ldots,z_Z]$ be a vector of length $Z$, then $L_1(\textbf{z})=\sum_{i=1}^Z |z_i|$ is the $L_1$ norm.

\section{VLC components and channel Model}\label{sec:Channel}
In VLC, unlike RF systems, data is conveyed on the intensity of the emitted light from the LEDs, therefore, frequency and phase modulations cannot be applied. Moreover, due to the characteristics of intensity modulation, transmitted signals must be positive and real valued. Also, to ensure that the LED is functioning in its dynamic range, the transmitted peak power should not exceed a particular constant value. In this section, different components of VLC systems are presented in addition to VLC channel modeling.

 %Then, we present the state-of-the-art literature on  MIMO in VLC, considering both single-user and multi-user cases. All used acronyms and abbreviations are summarized in Table \ref{Table1}.        
%U.

\begin{figure}[t]
%\centering
\centerline{\includegraphics[width=3.75in]{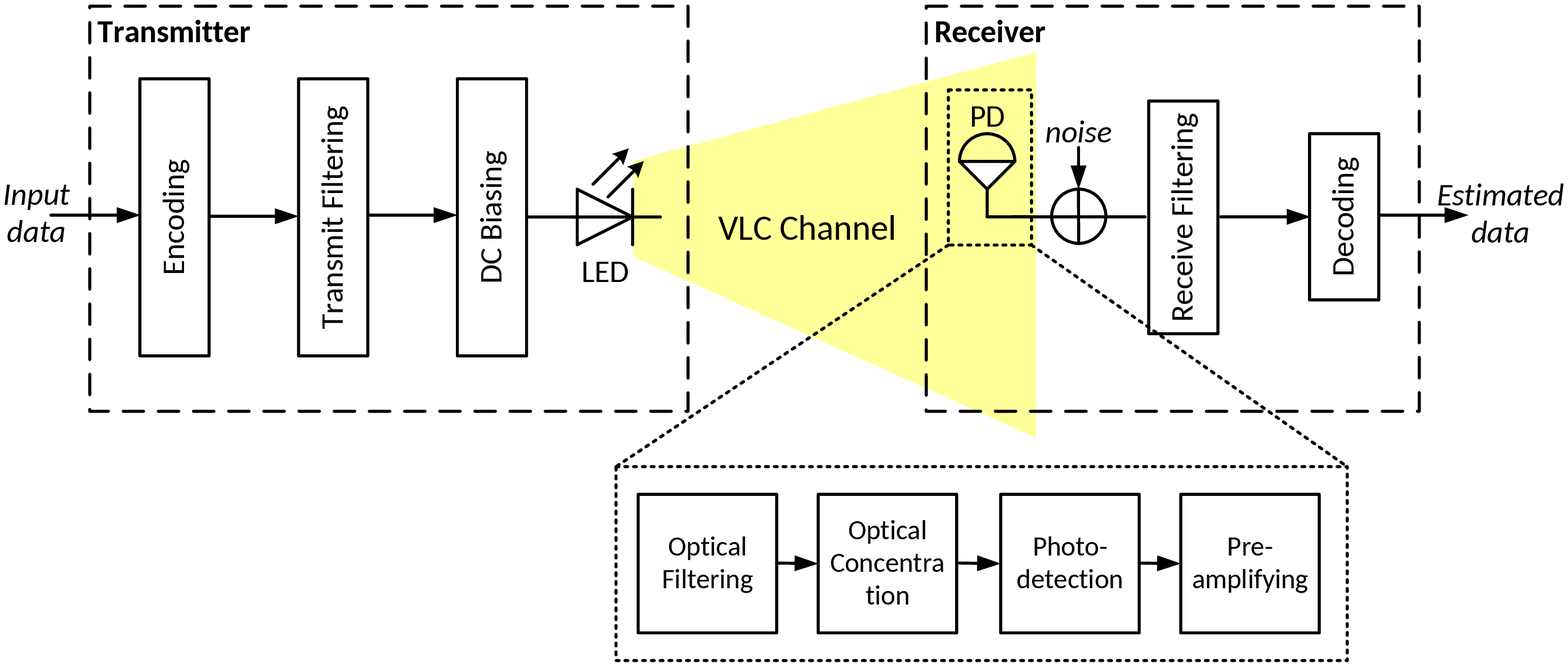}}
\caption{VLC transceiver components.}
\label{Fig:VLCComp}
\end{figure}

%\subsection{VLC Components} 
Fig. \ref{Fig:VLCComp} illustrates the basic VLC transceiver components. At the transmitter, an illuminating device is utilized for data modulation through IM/DD. There are a variety of light sources that are available for optical communication, but the commonly used ones are LEDs and laser diodes. Particularly, LEDs are the most popular illuminating devices, due to their low fabrication cost. They are composed of solid-state semiconductor devices that produce spontaneous optical radiation when subjected to a voltage bias across the P-N junction \cite{Sze2007}. The direct current (DC) bias excites the electrons resulting in released energy in the form of photons. In most buildings, white LEDs are preferred since objects seen under white LEDs have similar colors to when seen under natural light. Two common designs are considered for white LEDs. In the first design, a blue LED with a yellow phosphor layer is utilized, while in the second design, red, green, and blue (RGB) LEDs are combined together. The first method is more popular, due to its simplicity and low implementation cost. However, it suffers from limited modulation bandwidth due to the intrinsic properties of the phosphor coating. On the other hand, RGB LEDs are more suitable for color shift keying modulation, enabling higher achievable data rates \cite{Monteiro2014}. %\textcolor{blue}{LEDs offer many advantages over conventional illuminating devices, such as florescent, incandescent, and bulbs, which include longer operational life time, energy-efficiency (reaching up to 80\% compared to conventional devices), very low radiation heat, operation in extreme temperatures, simple directing and dimming properties, and, lastly, very high switching rate .}

It is recalled that VLC receivers comprise photo detectors (PDs),  also known as non-imaging receivers or  imaging (camera) sensors. These are used in order 
 to convert incident light power into electrical current proportional to light intensity.  A typical VLC receiver consists of an optical filter,  optical concentrator,  PDs and  pre-amplifier. The optical filter eliminates interference from  ambient light sources, while the optical concentrator enlarges the effective reception area of the PD without increasing its physical size. The optical concentrator is characterized by three parameters, i.e., field of view (FoV), refractive index, and radius. In order to increase the achievable diversity gain of an optical communication link, multiple receiving units can be deployed with different orientations, optical filters, and concentrators. However, such deployment comes at the expense of additional receiver size and complexity. To address this issue, an imaging sensor with a single wide FoV concentrator can be used to create multiple images of the received signals. Imaging sensors consist of an array of PDs that are integrated with the same circuit. It is worth noting that the required large number of PDs to capture high resolution photos renders them energy inefficient. It is noted that the PDs area of a  VLC system is much larger than the corresponding wavelength. Consequently, the multipath fading in an indoor VLC environment does not occur \cite{Dai2015,Marshoud2016}. Nevertheless, indoor optical links suffer from dispersion, modeled as linear baseband impulse response. Also, the indoor optical wireless channels can be assumed quasi-static, due to the relatively low mobility of users and connected objects in indoor environments. 

%Although having a large detector area have the advantage of providing spatial diversity, it usually comes at the expense of increased noise and junction capacitance, thus reducing the bandwidth of the receiver. Hence, a concentrator should be used before the photo detector to increase the overall collective area.

\begin{figure}[t]
\centering
\centerline{\includegraphics[width=3.5in]{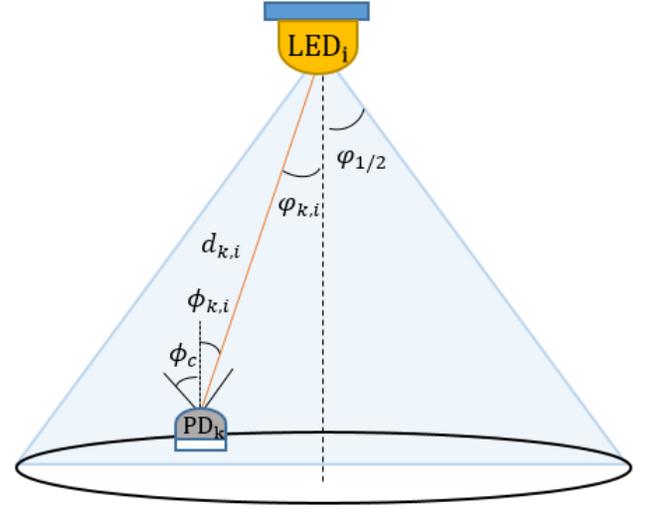}}
\caption{VLC channel model (link between LED $i$ and PD $k$).}
\label{Fig:channel1}
\end{figure}

Typically, the channel of a VLC link can be modeled as follows: With the non-line-of-sight components neglected in front of stronger line-of-sight (LoS) ones, the DC channel gain from the $i^{\rm th}$ LED to the $k^{\rm th}$ PD can be expressed by {\cite{Komine2004}}
\begin{equation}
\label{eq:channel}
\small 
 h_{k,i}= \left\{\begin{matrix*}[l]
\frac{A_{k}}{d^2_{k,i}} R_{o}(\varphi_{k,i}) T_{s}(\phi_{k,i}) g(\phi_{k,i}) \cos(\phi_{k,i}),  & 0 \leqslant \phi_{k,i} \leqslant \phi_{c}\\
0, & \text{otherwise,}
\end{matrix*}\right.
\end{equation}
where $A_{k}$ denotes the PD area, $d_{k,i}$ is the distance between the $i^{\rm th}$ LED and $k^{\rm th}$ PD, $\varphi_{k,i}$ is the transmission angle from the $i^{\rm th}$ LED to the $k^{\rm th}$ PD, $\phi_{k,i}$ denotes the incident angle with respect to the receiver, and $\phi_{c}$ is the FoV of the PD. These angles are well-illustrated in Fig. \ref{Fig:channel1}. Moreover, $T_{s}(\phi_{k,i})$ is the gain of the optical filter, and $g(\phi_{k,i} )$ is the gain of the optical concentrator, expressed as 

\begin{equation}
\label{eq:g}
g(\phi_{k,i})= \left\{\begin{matrix*}[l]
\frac{n^2}{\sin^2({\phi_{c}})},  & 0 \leqslant \phi_{k,i} \leqslant \phi_{c}\\
0, &  \phi_{k,i}>\phi_{c}.
\end{matrix*}\right.
\end{equation}
Here, $n$ is the refractive index and $R_{o}(\varphi_{k,i})$ is the Lambertian radiant intensity given by
\begin{equation}
\label{eq:R}
R_{o}(\varphi_{k,i})= \frac{m+1}{2\pi} \left(\cos(\varphi_{k,i})\right)^m
\end{equation}
with $m$ denoting the order of the Lambertian emission, namely \begin{equation}
\label{eq:m}
m = \frac{\ln{(2)}}{\ln\left({\cos(\varphi_{1/2})}\right)} 
\end{equation}
where $\varphi_{1/2}$ is the LED semi-angle at half power. For a typical VLC link, the received noise at the $k^{\rm th}$ PD can be modeled as a Gaussian random variable with zero mean and variance 
\begin{equation}
\label{eq:noise}
\sigma_k^{2} = \sigma^2_{k,\rm{sh}} + \sigma^2_{k,\rm{th}}
\end{equation}
where $\sigma^2_{k,\rm{sh}}$ and $\sigma^2_{k,\rm{th}}$ are the variances of the shot and thermal noises at the $k^{\rm th}$ PD, respectively. The shot noise is caused by the high rate of the physical photo-electronic conversion process, whose variance can be written as
\begin{equation}
\label{eq:noise2}
\sigma^{2}_{k,\rm{sh}} = 2qB\left(\zeta_k h_{k,i}x_{i}+I_{\rm{bg}}I_{2}\right) 
\end{equation}        
where $q$ represents the electronic charge, while $\zeta_k$ denotes the detector responsivity. Also, $x_{i}$ is the transmitted signal by the $i^{\rm th}$ LED, $B$ is the corresponding bandwidth, $I_{\rm{bg}}$ is the background current, and $I_{2}$ denotes the noise bandwidth factor. On the other hand, the thermal noise results from the transimpedance receiver circuitry and its variance at the $k^{\rm th}$ PD is given by
\begin{equation}
\label{eq:noise3}
\sigma^{2}_{k,\rm{th}} = \frac{8\pi K T_{k}}{G}\eta A_{k} I_{2} B^{2} + \frac {16 \pi^{2} K T_{k} \gamma}{g_{m}} \eta^{2} A^2_{k} I_{3} B^{3}
\end{equation}
where $K$ is the Boltzmann's constant, $T_{k}$ is the absolute temperature, $G$ is the open-loop voltage gain, $A_k$ is the PD area, $\eta$ is the PD's fixed capacitance per unit area, $\gamma$ is the field-effect transistor (FET) channel noise factor, $g$ is the FET transconductance, and $I_{3}$  = 0.0868 \cite{Komine2004}.
Modern infrastructures are commonly equipped with LED fixtures or arrays. A single fixture is composed of $Q$ LEDs, and may be viewed as a single VLC source,\footnote{In the remaining of this paper, we interchangeably designate by LED a fixture of LEDs.} with the DC channel gain given by
\begin{equation}
\label{eq:fixture}
\small 
h_{k,j}=\left\{
\begin{array}{ll}
A_k \sum \limits_{i=1}^Q d_{k,j,i}^{-2} R_o(\varphi_{k,j,i}) T_s(\phi_{k,j,i}) g(\phi_{k,j,i}) \cos(\phi_{k,j,i}),  \\
\;\;\;\;\;\;\;\;\;\;\;\;\;\;\;\;\;\;\;\;\;\;\;\;\;\;\;\;\;\;\;\;\;\;\;\;\;\;\;\;\;\;\;\;\;\;\;\;\;\;\;\;\;\;\;\;\;  0 \leq \phi_{k,j,i} \leq \phi_c \\
0, \; \mbox{otherwise}\end{array}
\right.
\end{equation}
where $d_{k,j,i}$ and $\varphi_{k,j,i}$ denote the respective distance and transmission angle between the $i^{\rm th}$ LED in the $j^{\rm th}$ fixture and the $k^{\rm th}$ PD, and $\phi_{k,j,i}$ is the incident angle with respect to the receiver. Since the separation between LEDs in the same fixture is negligible compared to the distance between the fixture and the $k^{\rm th}$ PD, then distances and angles implicating index $i$ can be assumed approximately the same for all LEDs. Hence, the channel gain from the $j^{\rm th}$ fixture to the $k^{\rm th}$ PD can be given by
\begin{equation}
\label{eq:fixture_app}
\small 
h_{k,j}\approx \left\{
\begin{array}{ll}
Q \; h_{k,i},  & 0 \leq \phi_{k,j,i} \leq \phi_c, \; \forall i  \\
0,  & \mbox{otherwise.}\end{array}
\right.
\end{equation}

In the following sections, we provide an in-depth study of the common MA schemes in the context of VLC.

%It was shown in RF systems that RSMA is a more generalized and powerful MA scheme that can support variety of network loads and users’ deployments. Therefore, a comparison between RSMA and different MA schemes will be also presented.

\section {\textcolor{black}{NOMA} for VLC} \label{sec:MA}

\subsection{Background}
%\textcolor{red}{The continuous increase of the number of connected nodes into the network requires the development of novel and efficient multiple access schemes, in order to facilitate the deployment of practical optical systems.} 
Inspired by promising multiplexing gains achieved by conventional MA techniques, which were developed for RF systems, optical MA have received great attention recently. To this end, conventional OMA schemes such as OFDMA and TDMA have been extensively studied in the context of VLC, in which users are allocated orthogonal  frequency/time resources. In the same context, several optical OFDM-based MA techniques were proposed, such as DC biased optical OFDM, asymmetrically clipped OFDM, asymmetrically clipped DC-biased optical OFDM, fast-OFDM and, polar-OFDM. However,  OFDM suffers from  high peak-to-average power ratio, which is difficult to overcome in VLC systems due to the non-linearity of LEDs \cite{Armstrong2009,Mesleh2011_2,Giaco2012,Dissan2013,Elgala2015}. OMA schemes efficiently mitigate interference among users’ signals by allocating orthogonal resources. However, the number of served users is limited and
cannot exceed the number of available orthogonal resources. This concern is also true for VLC systems.

Motivated by the above, researchers have recently focused on the design of novel NOMA techniques as a promising candidate to enhance spectral efficiency in 5G and beyond networks \cite{Ding2014}. The key principle is to allow different users to share the same frequency resources simultaneously at the expense of multi-user interference (MUI). To perform multi-user detection (MUD), different users are assigned distinct power levels based on their  channels gain, which is referred to as P-NOMA, or different spreading sequence, known as C-NOMA \cite{Ding2014,Ali2017,Wei2018}. \textcolor{black}{In RF systems, P-NOMA was considered as a candidate for the downlink communication of various standardizations activities  such as the 3rd generation partnership project (3GPP) standard LTE Release 13 \cite{Dobre}. Furthermore, P-NOMA has been envisioned as a key solution in 5G mobile systems \cite{Dai2015}.  On the other hand, both P-NOMA and C-NOMA have been considered for the uplink communication, in order to serve a larger number of users. }
In downlink P-NOMA systems, superposition coding (SC) at the base station and SIC at the receiver are utilized to transmit and detect the intended user's signal by eliminating users' signals with higher power levels, respectively. \textcolor{black}{While, in the uplink P-NOMA systems, the transmitted power is limited by the end users, the transmitted power of each individual user should be carefully adjusted such that the user with better channel gain will have more power contribution in the received signal. At the base station, the user with the best channel gain is decoded first. Then, a subsequent SIC is performed in order to decode the messages of the weaker users, which is the opposite of the downlink P-NOMA \cite{Dai2015,Dobre}}.

\textcolor{black}{In  C-NOMA, users are multiplexed in the code domain, in which each user is assigned a different code. Unlike the conventional code-division multiple access (CDMA), where dense spreading sequences are used, C-NOMA utilizes non-orthogonal sequences with low cross correlation or sparse spreading sequences to efficiently reduce the inter-user interference \cite{Shoreh,Chen_CDMA,Dai2018,Lou2018}, and hence, enhance the overall system performance. Specifically, optimal performance can be achieved in C-NOMA based VLC systems by exploiting optical code sequences \cite{Rashidi2013}. At the receiver, multi-user detection can then be realized by adopting message passing algorithms (MPA). It is noted that different
 versions of C-NOMA have been developed, such as low-density spreading (LDS)-CDMA \cite{Razavi2011}, low-density spreading (LDS)-OFDM \cite{Imari2011}, and sparse code multiple access (SCMA) \cite{cai}. LDS-CDMA utilizes low density spreading sequences in order to reduce the interference on each chip compared to the traditional CDMA. On the other hand, LDS-OFDM can be thought as a combination of both  OFDM and LDS-CDMA, where the resulted chips from the implementation of LDS-CDMA are transmitted over a set of sub-carriers. Finally, SCMA is a form of LDS-CDMA, where the information bits can be directly mapped to a distinct sparse codeword. Yet, although C-NOMA has a potential to enhance spectral efficiency, it requires additional bandwidth, challenging codebook design and is not easily applicable to the current systems compared to P-NOMA, which  has a simple implementation in the existing networks. For theses reasons, most of the research on NOMA systems have extensively considered the performance of P-NOMA  \cite{Tao,cai2018,Dai2015}. 
In VLC systems, the research on C-NOMA  is limited only to  \cite{Lou2018}. Therefore, it requires further investigation in the context of VLC.} \textcolor{black}{It is also worth noting that uplink VLC is impractical due to the power limitations of portable devices and the unpleasant radiance produced by end users. So, it is expected that the current VLC technologies will rely on RF or infrared in the uplink communications \cite{Ahmadi2018,Ding2019,alresheedi2017}.  Consequently, most of the research efforts on NOMA in VLC systems have focused on the downlink scenario.} A basic system model for two-user downlink P-NOMA in VLC systems is illustrated in Fig. \ref{Fig:PDNOMA}.

\textcolor{black}{For indoor VLC systems, P-NOMA is preferred for several reasons \cite{Marshoud_WCL}. First, P-NOMA depends on the channel state information, which can be readily available in VLC systems. Second, P-NOMA achieves the best performance in the high SNR regime, which is a common SNR  regime in VLC channels \cite{Ghassem2019}. Third, P-NOMA performs best when users’ channels are distinct. In VLC systems, the underlying symmetry issue of the channels has been addressed in  \cite{Marshoud2017}, where the authors proposed reducing channels' symmetry by an adaptive tuning of the semi-angle of the LEDs and the receiver’s FoV, as well as in \cite{Wang2018}, where advance receiver structures were considered. Finally, P-NOMA can be easily integrated with various technologies, such as MIMO and cooperative networks \cite{cai}. All these reasons motivated an important study focus on P-NOMA based VLC systems. }
\subsection{Superposition Coding and Successive Interference Cancellation}

\begin{figure}[t]

\centerline{\includegraphics[width=3.5in]{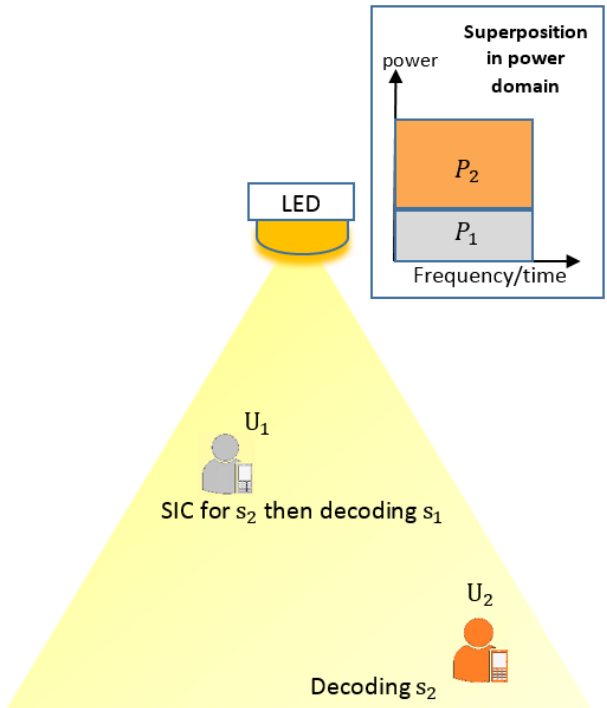}}
\caption{Scenario of two users P-NOMA in VLC.}
\label{Fig:PDNOMA}
\end{figure}

SC was first introduced in 1972 by Cover \cite{cover} as a method to transmit different signals to several receivers through a single source. To make SC practical, the transmitter encodes the data of two users as a two-layer single signal. Then, one receiver recovers the messages of the two layers, while the other recovers only one message from one-layer and ignores the message of the second layer. 
%SC is similar in concept to many practical situations, \textcolor{green}{such as giving a talk to an audience of different backgrounds. The speaker would manage the talk in a way to guarantee minimum information understanding by the weak-background audience, while ensuring that more extensive information is transmitted to the strong-background audience.} 
%This scenario is an example of broadcast of a superposed talk delivered by a speaker.
SC is realized  by allocating higher power coefficients  to users with the weakest channel conditions. On the other hand, users with strongest channel gains are assigned the lowest power levels\cite{Dobre}. For instance, in Fig. \ref{Fig:PDNOMA}, the second user $U_2$ is allocated fraction $P_2$ of the total power, and the first user $U_1$ is allocated $P_1$, such that $P_1<P_2$.
%The same principle can be generalized to any number of users in the system.
%\textcolor{red}{ Cover proposed  SIC technique that exploits the difference in the signals power levels in order to decode the superposed signals successively\cite{cover,Miridakis}. going back to the two users in Fig. \ref{Fig:PDNOMA}, 
Since the weaker user $U_2$ is allocated higher power, it can directly decode its signal while treating the signals of the other users as noise. On the other hand, using SIC, $U_1$ has to remove the interference by decoding $U_2$ signal, before detecting its own signal. 

%\textcolor{green}{Very short - No References?? If you can't make it longer, I don't see why it should be on a separate subsecition, why not merge it with the previous SC subsection and call it SC and SIC}.

\subsection{NOMA in VLC}

Several studies have considered the performance of different NOMA configurations in VLC systems. In \cite{Dai2018}, the symbol error rate (SER) of C-NOMA-based VLC system was investigated, where users demonstrated identical error rate performance at different locations. On the other hand, recent published papers have shown that in VLC systems, P-NOMA is an efficient MA scheme for the aforementioned reasons \cite{Lin2019,Marshoud2017,liu2019}.
%First, P-NOMA depends highly on the channel state information (CSI), which can be readily available in VLC scenarios. Second, P-NOMA achieves the best performance at high SNR, which is mostly the case for VLC channels. Third, NOMA performs efficiently when users’ channels are distinct. This can be achieved by changing the angle of the transmitting LED and the PD's FoV \cite{Marshoud2017}.
Furthermore, reported results showed that NOMA\footnote{In the remaining of the paper, P-NOMA is designated by NOMA.} outperforms OMA techniques, such as OFDMA and TDMA, in terms of system capacity and number of simultaneously served users, particularly in single-input single-output broadcast channels and for certain channel strength disparities among users \cite{Kizi2015,Yin2016,Yang2017,Shen2017}. The work in \cite{NOMA_random_receiver}  considered the performance of NOMA VLC system when users have random vertical orientations. In particular, the authors proposed users scheduling techniques and feedback mechanisms to boost the spectral efficiency. Moreover, a hybrid NOMA-OFDM system was investigated and shown to have superior performance in terms of the achievable rate over OMA-OFDM systems \cite{Chu2017,Fu2018}. Likewise,  for uplink NOMA-based VLC communications, joint detection was proposed in \cite{Guan} to decode the messages of multiple users.

Although NOMA is efficient for scenarios where the number of users is higher than the number of available orthogonal resources, its complexity grows rapidly and proportionally to the number of users, since the $k^{\rm th}$ user needs to decode the messages of the $k-1$ users before detecting its own signal. To address this issue, a simple approach is to group users into small clusters, such that users of the same cluster communicate using NOMA, while the different clusters are scheduled using an OMA technique. %It is worth mentioning that NOMA achieves interesting performances as long as users experience channels with different gains. 

Particularly, MIMO can be leveraged to provide additional gains for transmissions, which can be realized through precoded SC and hybrid SDMA-NOMA or OMA. In precoded SC, all users are sorted based on their effective precoded channel gains in a single cluster \cite{Wyner1974,Nguyen2017,Dobre2018}. On the contrary, in hybrid SDMA/NOMA/OMA, users are grouped in clusters separated by SDMA\footnote{SDMA is discussed in detail in Section \ref{sdma}.} \cite{Zeng2017_3}, where the users of a single cluster are served using NOMA. It was demonstrated in \cite{Zeng2017_3} and \cite{Zeng2017_2} that MIMO-NOMA outperforms MIMO-OMA in terms of sum rate and user fairness. However, despite the aforementioned advantages of MIMO-NOMA systems, they come at the expense of a complex transmitter design, where joint optimization of signals precoding/decoding orders is required for different users.  

As explained earlier, MIMO design can be realized by assuming multiple transmitting LEDs and multiple PDs at the receiver. Such a system cannot employ the same power allocation method designed for single transmitting LED NOMA VLC systems, such as gain ratio power allocation (GRPA) \cite{Marshoud2016}. Accordingly, several power allocation strategies have been proposed in the literature for MIMO-NOMA RF systems, e.g., hybrid precoding and post-detection \cite{Ding2016}, and signal alignment \cite{Ding2016_2,dobri2020}. However, their counterpart in the MIMO-NOMA VLC is almost non-existent. To our knowledge, only Chen \textit{et al.} investigated  NOMA-based MIMO VLC systems and proposed a power allocation strategy, called normalized gain difference power allocation (NGDPA) \cite{Chen2018}. The reported results for NGDPA illustrated a sum rate improvement of 29.1\% compared to GRPA.

Multi-cell NOMA in the context of VLC has also not been well-investigated in the literature. In \cite{Zhang2017}, Zhang \textit{et al.} proposed a user grouping scheme based on users locations to minimize the interference caused by multi-cell deployment. Further, the authors in \cite{Rajput2019} proposed a joint NOMA transmission scheme to serve users in overlapping regions of different cells. In \cite{Shi2019}, Shi \textit{et al.} investigated the use of offset quadrature amplitude modulation (OQAM)/OFDM-NOMA modulation in a multi-cell VLC system. \textcolor{black}{Although NOMA schemes outperform OMA-based VLC systems in terms of spectral efficiency, the multiplexing gain of NOMA is highly affected by channel symmetry. In the context of VLC, channel symmetry is a major challenge as the communication is usually due to the LoS scenario \cite{Ahmadi2018,Marshoud_WCL}. Indeed for small cell design, where the number of users is considered relatively large, it is inefficient to multiplex all users using NOMA, as this may lead to an increased complexity and unsuccessful SIC operation. 
Therefore, hybrid NOMA/OMA schemes are promising solutions to realize a trade-off between multiplexing gain, computational complexity, and error propagation. In hybrid NOMA/OMA systems, users are split into multiple groups, where users within the same group employ NOMA, while different groups are multiplexed using OMA. User pairing and grouping represents key challenge in hybrid NOMA/OMA systems that requires sophisticated algorithms to have the full potentials of NOMA. However in VLC systems, as users mobility is low, the variation in the channel conditions is relatively small, hence, user pairing and grouping requires less complicated algorithms compared to RF systems. The effect of user paring and grouping has been extensively studied in RF systems \cite{Benjebbovu2013,Sedaghat2018,Abbasi2016,Zhu2019,He2016}, but, its application in VLC systems requires further investigations.    
For instance, in \cite{Yin2016, Almohimmah2018}, the authors adopted channel gain-based pairing strategy that aims to maximize the system's throughput. This strategy relies on selecting two users with the most distinctive channel gains to perform NOMA. However, their approach causes high interference to users with correlated channels. In \cite{Yapici2019}, the authors proposed individual and group-based NOMA users pairing in a VLC system and showed that near-optimal sum-rate performance can be achieved.
Moreover, user grouping based on users' locations was proposed in \cite{Zhang2017} in order to reduce the interference in VLC multi-cell networks, where users in each group are served by only one access point using NOMA. %Also, authors of \cite{zhou_wong2018} proposed a users pairing strategy, in which the coverage area is divided into far and near regions and each user in the far region is paired with one user in the near region. However, when the number of users is unbalanced between the far and near regions, this strategy results in an increased number of orphan users.
On the other hand, the authors in \cite{Abumarshoud2019} proposed a hybrid OMA/NOMA scheme for attocellular VLC based on a smart transmitter to select dynamically the adequate MA technique according to the environment conditions.  Finally, in \cite{Janjua2020}, the authors proposed an efficient user paring for the cases of having odd and even number of users. First users are ordered in ascending order based on the channel strength, then, they are either grouped into two or three groups depending on whether the number of the uses is even or odd.  Then pairing is performed by choosing one user from each group starting from the users with the lowest channel gain.  Reducing  channels' correlation for the end users is another method of enhancing the performance of NOMA system. In \cite{Marshoud2017}, the authors proposed an adaptive adjustment of  the  semi-angle  of  the  LEDs  and  the  FoV  of  the  PD in order to create dissimilar channels. Additionally, the use of different advanced receiver structures to reduce channels correlation is a potential solution that may be further investigated \cite{Wang2018}. Nevertheless, the enhancement of NOMA performance through reducing channels symmetry need to be further explored in the context of VLC.  } Finally, it is noted that NOMA based hybrid VLC/RF systems have been proposed as acceptable solutions that compensate for the limitation of VLC systems, particularly in uplink communication scenarios \cite{Papanikolaou2019,Papanikolaou_GLOBECOM,Zhou2018}. In the same context, hybrid wavelength division multiplexing (WDM)-NOMA has been proposed in \cite{Aljohani}, where multi-color LEDs are used to allow simultaneous transmissions at different wavelengths. 
Relevant work on NOMA in VLC is summarized in Table \ref{TableI}.

\begin{table*}%{|p{30pt}|p{90pt}|p{75pt}|p{100pt}|p{140pt}|}%[t]
%\centering
\caption{VLC Related Work on NOMA.}
\label{TableI}
\centering
\footnotesize
\begin{tabular}{|p{30pt}|p{120pt}|p{120pt}|p{180pt}|}
\hline
{\textbf{Reference}} & \makecell{\textbf{System Model}} & \makecell{\textbf{Objective}} &  \makecell{\textbf{Findings}}  \\
\hline 
\hline

\makecell{\cite{Dai2018}} & \makecell{Single-cell downlink C-NOMA} & \makecell{Analysis and evaluation of the SER} &  \makecell{Using adequate power allocation, users at\\ different locations achieve almost an identical SER } \\ \hline

\makecell{\cite{Marshoud2017}} & \makecell{Single-cell downlink NOMA} & \makecell{Derivation of closed-form expression\\ for the bit error rate (BER)} &  \makecell{Closed-form expressions match simulation results} \\ \hline

\makecell{\cite{Yin2016}} & \makecell{Single-cell  downlink NOMA} & \makecell{ Derivation of closed-form expressions\\ for the coverage probability\\ and ergodic sum rate } &  \makecell{ NOMA outperforms conventional OMA scheme} \\ \hline

\makecell{\cite{NOMA_random_receiver}} & \makecell{Single-cell downlink NOMA} & \makecell{Derivation of closed-form expressions\\ for the sum rate and outage\\ probability } &  \makecell{Analytical results agree with simulation results \\ and near-optimal sum rate is achieved using \\ a limited feedback scheme} \\ \hline

\makecell{\cite{Chu2017},\cite{Fu2018}} & \makecell{Single-cell downlink NOMA-OFDM} & \makecell{Maximization of the sum rate} &  \makecell{NOMA-OFDM is superior to OMA-OFDM system,\\ in terms of achievable data rate} \\ \hline

\makecell{\cite{Guan}} & \makecell{Single-cell  uplink NOMA} & \makecell{Evaluation of BER based on phase \\ pre-distorted joint detection } &  \makecell{Improved BER performance compared to \\NOMA based on SIC for different power values } \\ \hline

\makecell{\cite{Chen2018}} & \makecell{MIMO-NOMA} & \makecell{Maximization of the sum rate} &  \makecell{NGDPA improves the sum rate \\ performance compared to GRPA } \\ \hline

\makecell{\cite{Zhang2017}} & \makecell{ Downlink MU-multi-cell NOMA} & \makecell{Maximization of the sum rate\\ and  max-min rate criterion} &  \makecell{User grouping and power allocation optimized, \\hence achieving higher sum user rate than OMA
} \\ \hline

\makecell{\cite{Rajput2019}} & \makecell{Downlink MU-multi-cell NOMA} & \makecell{Maximization of the sum rate } &  \makecell{Joint transmission (JT) NOMA achieves 
higher sum\\ rates compared to the frequency reuse factor-2 NOMA} \\ \hline

\makecell{\cite{Shi2019}} & \makecell{ Downlink  MU-multi-cell\\ OQAM/OFDM-NOMA} & \makecell{ Evaluation of spectral efficiency,  \\ BER, and error vector magnitude} &  \makecell{Proposed scheme outperforms OFDM-NOMA\\ and is more robust to inter-cell interference
} \\ \hline

\makecell{\cite{Abumarshoud2019}} & \makecell{Multi-cell hybrid OMA/NOMA} & \makecell{Evaluation of sum rate, outage\\ probability, and fairness performances} &  \makecell{Dynamically selecting the adequate MA technique\\ achieves better performances that static configuration} \\ \hline

\makecell{\cite{Papanikolaou2019,Papanikolaou_GLOBECOM}} & \makecell{Downlink hybrid VLC/RF} & \makecell{Maximization of the sum rate} &  \makecell{Optimal joint user grouping and power allocation \\based on game theory was proposed; this outperforms\\ the standard opportunistic scheme} \\ \hline

\makecell{\cite{Zhou2018}} & \makecell{Cooperative NOMA VLC/RF with \\simultaneous wireless information\\ and power transfer (SWIPT)} & \makecell{Derivation of closed-form expression \\for the outage probability} &  \makecell{A trade-off on rate splitting allows outage performance\\ balancing among users} \\ \hline

\makecell{\cite{Aljohani}} & \makecell{Hybrid WDM-NOMA} & \makecell{Maximization of the sum rate} &  {WDM-NOMA outperforms NOMA in terms of sum rate} \\ \hline

\end{tabular}
\end{table*}

\section {\textcolor{black}{SDMA} for VLC} \label{sdma}
\subsection{Background}
In recent building designs, it is common to have multiple illuminating LEDs in indoor spaces. In VLC systems, channel access can be realized either through multiple access channel (MAC) or broadcast channel (BC). Most research effort has focused on the analysis of downlink BC of MU-MIMO, with an emphasis on the data rate performance. Such systems experience interference when orthogonal frequency/time resources are limited. MUI is a common issue in MU-MIMO systems, which can be eliminated at the receiver using an efficient MUD technique \cite{Spencer2004,Spencer2004_2}. However, the implementation of MUD in VLC systems suffers from high complexity and energy inefficiency. Therefore, SDMA, which is based on data precoding at the transmitter, constitutes a promising alternative solution. 

%\textcolor{green}{ADD A TABLE ABOUT RELATED WORK SDMA-VLC}

\subsection{SDMA in VLC}
An early implementation of SDMA is based on block diagonalization (BD), a generalized form of channel inversion precoding \cite{Chen2013}. Although BD is a simple linear precoding technique, its application is limited to the scenario where the number of transmitting LEDs is larger than the total number of served users, i.e., overloaded regime. The authors in \cite{Hong2013} used BD precoding in downlink MU-MIMO VLC and showed that BD is constrained by the correlation of the involved wireless links.  Hence, a scheme that reduces this correlation was proposed, based on the adjustment of PDs' FoVs. Yu \textit{et al.} developed in \cite{Yu2013} linear zero-forcing (ZF) and ZF dirty paper coding (ZF-DPC) schemes in order to eliminate MUI and maximize the throughput or max-min fairness. However, in \cite{Ma2013}, the authors relaxed the ZF condition by applying the minimum mean squared error (MMSE) as a performance metric for the precoder design, in both perfect and imperfect CSI scenarios. In \cite{Li2015}, an optimal mean squared error (MSE) precoder was designed in order to minimize the BER, under per-LED power constraints. The transceiver design was later simplified by adopting a ZF precoder. The corresponding results  showed that the simplified scheme outperforms the conventional ZF precoder in terms of BER, while MSE achieves the best performance. Similar designs were proposed in \cite{Pham2017}, where an optimal ZF precoder was obtained using an iterative concave-convex procedure, aiming at maximizing the achievable per-user data rate. Then, the authors simplified the precoder design using the high signal-to-noise ratio (SNR) approximation. In \cite{Shen2016_2}, Shen \textit{et al.} proposed a different beamforming technique aiming at maximizing the sum rate of a virtual MIMO VLC system. Beamforming was designed using the sequential parametric convex approximation method, and it has been shown through simulations that it outperforms conventional ZF-based beamforming, particularly  for highly correlated VLC channels and low optical transmit power. Likewise, Marshoud \textit{et al.} developed in \cite{Marshoud_2018} an optical adaptive precoding for downlink MIMO VLC systems, under perfect and imperfect CSI. 
BER results showed that their scheme is more robust to imperfect CSI and channel correlation than conventional channel inversion precoding.
Authors of \cite{Wang2015} proposed precoding for an OFDM-based MU-MIMO VLC system, where precoding is applied at each sub-carrier, using ZF and MMSE techniques.
This led to the enhancement of the sum rate performance at high SNR and for the case of uncorrelated channels. In \cite{Zeng_2017}, the sum rate maximization problem was reformulated as a weighted MMSE (WMMSE) problem to jointly design the BD precoding and receive filter coefficients. A similar approach was also considered in \cite{Ying2015}. Finally, Adasme \textit{et al.} proposed in \cite{Adasme} a hybrid approach, called spatial TDMA (STDMA), where full connectivity is achieved
 by allowing simultaneous data rate transmission of multiple nodes within an optimized schedule.

The contribution in \cite{Ma2015,Ma2018} focused on precoding designs for coordinated multi-point (CoMP) MU-MIMO VLC systems. Through numerical analysis, the authors showed improvements in terms of signal-to-interference-plus-noise ratio (SINR) and weighted sum MSE (WSMSE), respectively. Additionally, Yin \textit{et al.} considered in \cite{Yin2015,Yin_GLOBECOM} different SDMA grouping algorithms to obtain a trade-off between the Jain’s fairness index and area spectral efficiency for a CoMP-VLC system through the utilization of linear ZF precoding.
The authors in \cite{Yangchen2018}  proposed a joint precoder and equalizer design based on interference alignment for MU multi-cell MIMO VLC systems under imperfect CSI. In \cite{Pham2019}, different levels of coordination/cooperation were considered using a ZF precoder. 

It is worth noting that SDMA can be also realized using an angle diversity transmitter (ADT), which consists of multiple directional narrow-beam LED elements. 
%is also commonly used as an optical transmitter for SDMA instead on multiple distributed LEDs. 
An ADT creates independent narrow-band beams (by reducing the FoVs of LEDs) towards spatially deployed users, while achieving the same coverage as a single wide-beam transmitter \cite{Carrut2000,Kim2014,Chen2015}. ADTs can also replace conventional single-element transmitters in multi-cell scenarios such that more power is directed towards each user, and hence, improving the communication's reliability \cite{Basnayaka}. 
%While SDMA in RF systems requires a complex beamforming  to create the narrow-band beams, it is simpler in VLC when ADTs are used, where narrow-band beams can be achieved by reducing the FoVs of LEDs. 
In order to avoid interference among  users, spatial separation needs to be implemented by adequately allocating transmit power to the beams. Subsequently, each receiver attempts to detect its signal by treating any interference as noise. In spite of the ADT's interference reduction potential, it requires a complex optical front-end to supply independent signals to multiple LED elements.\\
%Spatial grouping is essential in the absence of beamforming to reduce the interference due to the small distance between users in VLC. 
%At the transmitter side spatial separation between the users can be achieved by utilizing linear precoding schemes to allocate fractions of the total transmit power to different beams. Therefore, SIC is not required at the receiver side, and the interference from other users is treated as noise, allowing the intended user to detect its own signal. 
Compared to NOMA, SDMA simplifies the transmitter and receiver designs. However, it becomes inefficient as soon as the number of users exceeds the number of transmit LEDs, i.e., an overloaded scenario. It should be noted that the number of LEDs has to be more than or equal to the number of users in order to guarantee interference reduction. Moreover, since SDMA depends highly on the CSI at the transmitter (CSIT) in order to mitigate interference, its performance degrades with imperfect CSI \cite{Guanghan,Sifaou,Marshoud_2018}. Finally, due to the unique characteristics of IM/DD, which require signals to be real and unipolar, it is very difficult to pair orthogonal users together, as in RF systems. 
%Consequently, SDMA performance degrades when used in VLC systems. 
The accomplished work on SDMA in VLC is summarized in Table \ref{TableIII}.

%SDMA is highly dependent on pairing orthogonal users together, which is not the case in VLC systems in which channels have real values, yielding a degraded system's performance.

\begin{table*}%{|p{30pt}|p{90pt}|p{75pt}|p{100pt}|p{140pt}|}%[t]
%\centering
\caption{VLC Related Work on SDMA.}
\label{TableIII}
\centering
\footnotesize
\begin{tabular}{|p{30pt}|p{120pt}|p{120pt}|p{180pt}|}
\hline
{\textbf{Reference}} & \makecell{\textbf{System Model}} & \makecell{\textbf{Objective}} &  \makecell{\textbf{Findings}}  \\
\hline 
\hline

\makecell{\cite{Chen2013,Hong2013}} & \makecell{MU-MISO with BD precoding} & \makecell{Evaluation of the BER  } &  \makecell{With enough transmit power, a data rate of 100 Mbps \\is achieved for BER=$10^{-6}$} \\ \hline

\makecell{\cite{Yu2013}} & \makecell{MU-MISO with ZF\\ or ZF-DPC precoding} & \makecell{ Maximization of the throughput \\and max-min fairness} &  \makecell{ZF-DPC outperforms linear ZF, in particular when\\ users are close to each other} \\ \hline

\makecell{\cite{Ma2013}} & \makecell{MU-MISO with MMSE precoding} & \makecell{Evaluation of the optimal linear\\ MMSE precoder under perfect \\ and imperfect CSIT} &  \makecell{Linear MMSE precoding is able to separate \\the broadcast signals at the VLC
receivers} \\ \hline

\makecell{\cite{Li2015}} & \makecell{MU-MISO with MMSE/ZF precoding} & \makecell{Minimization of the MSE\\ and evaluation of the BER } &  \makecell{MMSE precoding achieves beast results, \\while proposed simplified ZF approaches MMSE \\ performance for a small number of (or dispersed) users 
} \\ \hline

\makecell{\cite{Pham2017}} & \makecell{MU-MISO with ZF precoding } & \makecell{Maximization of the sum rate \\ and max-min fairness} &  \makecell{The generalized-inverse ZF design achieves better\\
 performance than the pseudo-inverse ZF design, \\in particular for high SNRs} \\ \hline

\makecell{\cite{Shen2016_2}} & \makecell{MU-MISO with ZF precoding} & \makecell{Maximization of the sum rate } &  \makecell{The proposed approach does not restrict the co-channel\\ interference to zero, and thus, achieves a higher sum\\ rate than conventional ZF techniques
} \\ \hline

\makecell{\cite{Marshoud_2018}} & \makecell{MU-MISO with adaptive precoding } & \makecell{Derivation of closed-form expression\\ and evaluation of BER under perfect\\ and imperfect CSIT} &  \makecell{Adaptive precoding provides
significant performance \\enhancement compared to conventional channel \\inversion precoding} \\ \hline

\makecell{\cite{Wang2015}} & \makecell{MU-MIMO OFDM with ZF/MMSE \\ precoding } & \makecell{Evaluation of the spectral efficiency } &  \makecell{Sub-carrier with higher index achieves a higher\\ spectral efficiency, particularly for highly correlated \\users, and MMSE outperforms ZF for low \\transmit power and close users} \\ \hline

\makecell{\cite{Zeng_2017}} & \makecell{MU-MISO with BD precoding} & \makecell{Maximization of the sum rate\\ with imperfect CSIT} &  \makecell{Robust precoding is designed using BD and WMMSE \\ to suppress MUI and channel estimation errors} \\ \hline

\makecell{\cite{Ying2015}} & \makecell{MU-MIMO with joint \\ MMSE precoding and equalizing} & \makecell{Minimization of the MSE \\and evaluation of the BER\\ in presence of CSIT errors} &  \makecell{Proposed joint optimization method demonstrates BER \\improvements when experiencing imperfect CSIT} \\ \hline

\makecell{\cite{Adasme}} & \makecell{MU-STDMA} & \makecell{Minimization of the total scheduling\\ time and power consumption } &  \makecell{STDMA achieves full
connectivity, and the \\proposed greedy algorithm significantly reduces \\the processing time} \\ \hline

\makecell{\cite{Ma2015, Ma2018}} & \makecell{Multi-cell MU-MIMO CoMP \\with MMSE precoding} & \makecell{Minimization of the WSMSE} &  \makecell{Proposed approach realizes low-complexity interference\\ mitigation compared to CoMP JT} \\ \hline

\makecell{\cite{Yin2015,Yin_GLOBECOM}} & \makecell{CoMP SDMA with ZF precoding} & \makecell{Evaluation of the Jain’s fairness index\\ and of area spectral efficiency} &  \makecell{The proposed grouping algorithm achieves
better\\ area spectral efficiency-fairness trade-off compared\\ to existing benchmarks} \\ \hline

\makecell{\cite{Yangchen2018}} & \makecell{Multi-cell MU-MIMO joint\\ MMSE precoding and equalizing } & \makecell{Minimization of the MSE \\ and sum rate evaluation\\ with imperfect CSIT} &  \makecell{The joint design of the precoder and equalizer\\ efficiently reduces inter-user interference and\\ inter-cell interference, and achieves better performance\\ compared to existing MMSE and max-rate designs} \\ \hline

\makecell{\cite{Pham2019}} & \makecell{Multi-cell MU-MIMO CoMP\\ with ZF precoding } & \makecell{Maximization of the sum rate} &  \makecell{Partial cooperative precoding and coordinated precoding\\ outperform per-cell coordinated precoding when the \\number of users is not large compared to the\\ number of the LEDs or for high transmit power} \\ \hline

\makecell{\cite{Kim2014,Chen2015}} & \makecell{SDMA using ADTs} & \makecell{Evaluation of the throughput} & \makecell{The use of ADTs improves the performance of\\ multi-user systems, and optical SDMA outperforms \\optical TDMA in terms of throughput} \\ \hline

\makecell{\cite{Basnayaka}} & \makecell{Attocell SDMA downlink using ADT} & \makecell{Derivation of closed-form \\expression for the spectral efficiency} &  \makecell{inter-cell interference is mitigated and optical SDMA\\ outperforms optical TDMA} \\ \hline

\makecell{\cite{Sifaou}} & \makecell{MU-MIMO using ADTs} & \makecell{Maximization of the minimum SINR\\ and evaluation of the rate per-user\\with imprefect CSIT} &  \makecell{The proposed precoding and receiver design is robust \\to channel estimation errors and achieves significant \\gains compared to non-robust receiver design} \\ \hline

\end{tabular}
\end{table*}

\section {\textcolor{black}{RSMA} for VLC}

%\textcolor{green}{we summarized the main contributions on NOMA and SDMA when used within the VLC context.}
%we summarized NOMA and SDMA  techniques and  integration with VLC systems. 
Although NOMA realizes simultaneous transmission of a large number of users in overloaded scenarios and SDMA achieves spatial separation between users in underloaded scenarios, their performance is highly dependent on users' deployments, channel conditions, and  availability of CSIT. %\textcolor{green}{Moreover, MIMO-NOMA suffers from inefficient users' spatial separation when used in broadcast channels}  
%in a MIMO BC channels leads to an inefficient utilization of the spatial dimension since channels can not be ordered based on the channel gain as in the SISO case as they are no longer scalars.} I DONT AGREE WITH THIS, BECAUSE YOU CAN EVALUATE THE CHANNEL GAIN OF A MIMO CHANNEL USING THE FROBINIUS NORM
Therefore, a generalized configuration, which can optimize the utilization of resources for both overloaded and underloaded scenarios and provide more robustness to CSIT estimation errors, is of paramount importance. This was the main motivation behind the proposal of RSMA as a generalized scheme, where NOMA and SDMA can be considered as special cases.

\begin{figure*}[t]
\centering
\includegraphics[width=1\linewidth]{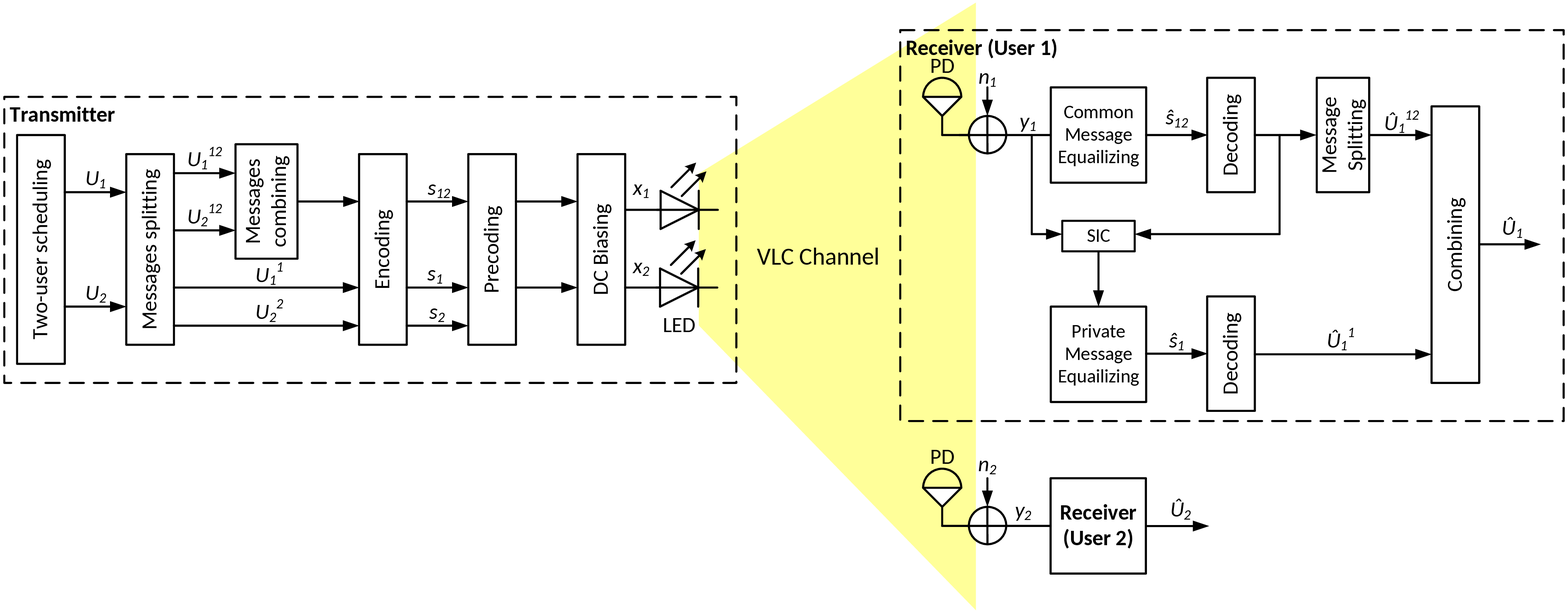}
\caption{RSMA-based two-user MISO VLC system.}
\label{Fig:2MISO}
\end{figure*}

\subsection {Background}

The basic concept of rate-splitting was first introduced in \cite{Te_Han} for a two users single-input single-output BC scenario. In \cite{Mao2018,Clerckx2016}, RSMA was proposed as a powerful and generalized MA technique for RF systems. It was demonstrated that RSMA could potentially offer tactile improvements as a MA technique, by allowing wireless networks to efficiently serve multiple users with different capabilities in overloaded and underloaded scenarios. According to the key principle of RSMA, which relies on the implementation of  linear precoding at the transmitter and SIC at the receiver, it has been shown that this MA scheme is capable of bridging the gap between NOMA and SDMA techniques. 
%At the receiver, RSMA allows decoding a part of the interfering signals and treating the other part as a noise. To do so, RSMA technique can be implemented as the following. 
In RSMA, users' messages are split into common and private parts at the transmitter. Then, a combiner is used to multiplex the common parts of all users and encode them into a single common stream. Meanwhile, the private parts are encoded separately into multiple private streams. Subsequently, a linear precoder is used to mitigate MUI. Finally, all precoded streams are superimposed on the same signal and sent over a VLC BC channel. At each user, the common stream is decoded and the user's intended data is extracted. Then, interference introduced by the common stream is eliminated using  SIC, as in NOMA. Subsequently, the private part of each user message can be decoded, while treating the private parts of other users messages as noise, as in SDMA. This mechanism is illustrated in (Fig. 1, \cite{Mao2018}).\\
RSMA depends mainly on the splitting design of  messages and  power allocation strategies between  common and private parts of users' messages. Extensive research efforts have been devoted to the investigation of these issues in order to improve the  efficiency of RSMA  %\cite{Mao2018,Dai2016,Dai2017,Joudeh2017,Papaz2017,Abdelhamid,Joudeh2016,Joudeh_ISIT,Mao_letter,Mao_2019} 
in the context of RF. In \cite{Mao2018}, the authors provided an analytical framework to study the performance of RSMA in MU-MISO BC channels. The reported results proved that RSMA outperforms NOMA and SDMA in terms of sum rate for different users' setups. Dai \textit{et al.} investigated in \cite{Dai2016} RSMA with massive MIMO and imperfect CSIT. They proposed a hierarchical-rate-splitting (HRS) framework where two different types of common messages are defined, which can be decoded by either all users or by a subset of them. Then, the associated sum rate performance was investigated in order to adjust the precoders of common messages. Numerical results illustrated the superiority of HRS compared to conventional techniques such as TDMA and BC with user scheduling. This work was extended in \cite{Dai2017} to a MU millimeter wave (mmWave) case, where CSIT is either statistical or quantized. Similar to \cite{Dai2016}, Joudeh \textit{et al.} proposed in \cite{Joudeh2017} a hybrid RSMA messages precoding in order to achieve max-min fairness amongst multiple co-channel multicast groups. The superiority of their approach is proved through degree-of-freedom analysis and simulation results. The authors in \cite{Papaz2017} evaluated the robustness of RSMA, in the presence of hardware impairments, such as phase distortion and thermal noises, and the availability of perfect/imperfect CSIT. In addition, Abdelhamid \textit{et al.} investigated in \cite{Abdelhamid} the use of channel inversion precoding for MU-MIMO RSMA system, where phase-shift-keying was the adopted modulation scheme. Results showed that RS combined with channel inversion has a significant sum rate improvement compared to RS with ZF or other MA schemes. Moreover, the authors in \cite{Yijie} incorporated RS with DPC to achieve the largest rate region for MISO BC with partial CSIT for different network loads and users' deployments. 
In \cite{Hao2015}, Hao \textit{et al.} proposed a practical scheme for private symbols encoding in RSMA using the conventional ZF beamforming. Then, they studied the sum rate performance for a two-user BC channel with limited CSI feedback.   
In \cite{Gui-Zhou}, the authors considered the trade-off between the spectral efficiency and energy efficiency for RSMA in multi-antenna BC channels. 
%The optimization problem for the SE/EE was a multi-objective function, hence, the author suggested two approaches namely, weighted-sum approach and weighted-power approach. 
It was shown that RSMA achieves a significant improvement in terms of spectral and energy efficiency. The use of RSMA in the downlink of a MISO SWIPT BC channel was investigated in \cite{Mao_swipt}. The sum rate performance of rate-splitting was evaluated and compared with other MA schemes. A further study of
RSMA in downlink CoMP JT networks was considered in \cite{Mao_JT_2019}, where results showed the superiority of RS in JT over SDMA- or NOMA-based JT. Also, in \cite{Zhang2019}, Zhang \textit{et al.} considered a cooperative rate splitting strategy based on the three-node relay channel, and demonstrated the enhanced performance of this scheme compared to cooperative-based NOMA. Similar results were reported in \cite{Yijie2019}, where the max-min fairness was used as a metric for a $K$-user MISO BC with user's relaying cooperative communication. 
The authors in \cite{Papazafeiropoulos} adopted an RS strategy to overcome the saturation occurred in multi-pair MIMO relay systems with imperfect CSIT. %It was found that by increasing the number of relay antennas or by decreasing the severity of self-interference, RS appears to be more robust.
The use of RSMA in cloud radio access network was considered in \cite{yu2019}. Yu \textit{et al.} proposed an enhanced RSMA technique that outperforms the original RSMA through careful grouping of common signals that are chosen using hierarchical clustering with inter-UEs dissimilarity metric, defined based on channel directions. Additionally, the superiority of RSMA over other MA schemes was investigated for satellite systems in \cite{Longfei}, where  users achieved max-min fairness for multi-beam satellite communications under CSIT uncertainty with minimum inter-beam interference. Finally, RSMA was considered in \cite{Rahmati} for cellular-connected drones, where the authors investigated the energy efficiency of RSMA and NOMA schemes in a mmWave downlink transmission scenario. % the design of the precoder was done taking into account the 3GPP antenna propagation patterns.

%\textcolor{green}{THIS SECTION NEEDS MORE REFERENCES, as a REVIEWER has SHOWN there are 35 references on RSMA from Clerckx only! You can be assisted by some of them cited in the other paper we're correcting.}

Despite the extensive research efforts on RSMA for different systems in the RF domain, its applicability in VLC systems has not yet been explored. Therefore, in this article, we provide preliminary results on the performance of RSMA in VLC systems, and compare its capacity gain with respect to existing VLC MA techniques. Furthermore, we give insights into the challenges and future research directions for RSMA-based VLC systems.

\subsection{RSMA-Based VLC}

%RSMA was initially proposed for RF communication systems in \cite{Elgamal2011}, as a method of achieving a new rate region for  two-user BC channels. 
The concept of RSMA was proposed to a multi-antenna BC channel in \cite{Mao2018} to bridge the gap between two extreme MA schemes, namely NOMA and SDMA. Mao \textit{et al.} showed that RSMA works best in the multiple-input case. In the VLC context, this can be realized using several transmitting LEDs to create a BC channel towards several users. 
Hence, in this survey, we analyze the performance of RSMA in a downlink MU-MISO BC VLC system. 
%Its benefits come from decoding part of the interference while treating the other part as a noise. 

%To recall, RSMA can be realized by splitting users' messages into common and private parts. Common parts form different users are encoded onto a single common stream, and is decoded by each individual user. On the other hand, the private parts of all users, which are encoded separately into multiple private streams, are only decoded by the intended user. 

\subsubsection {System Model} \label{sec:models}
For the sake of simplicity but without a loss of generality, we assume two transmitting LEDs that send messages to two single-PD users, as depicted in Fig. \ref{Fig:2MISO}. Messages $U_1$ and $U_{2}$ are intended to users 1 and 2, respectively. $U_1$ is divided into two parts: private part $U^1_{1}$ and common part $U^{12}_{1}$. Similarly, $U_2$ is divided into $U^2_{2}$ and $U^{12}_{2}$. Then, the two private messages, $U^{1}_{1}$ and $U^{2}_{2}$, are encoded into private streams $s_{1}$ and $s_{2}$, respectively. Then, from a common codebook, $U^{12}_{1}$ and $U^{12}_{2}$ are combined and encoded into one common stream $s_{12}$.
Without loss of generality, we assume that $s_{i}$ ($i \in \{1, 2, 12\}$) is randomly selected from a pulse amplitude modulation constellation with zero mean and normalized range $\{-1, 1\}$. Let $\textbf{s}=\left[s_1,s_2,s_{12}\right]^T$ be the transmitted symbols vector, with $\mathbb{E}(\textbf{s}\textbf{s}^{T})=\textbf{I}$. It is further assumed that the non-linear response of the LED is compensated through digital pre-disposition \cite{Stepniak2013}. To reduce MUI, a linear precoding matrix $\textbf{P}=\left[\textbf{p}_{1},\textbf{p}_{2},\textbf{p}_{12}\right]$ is considered, where $\textbf{p}_{i}=[p_{i,1} \, p_{i,2}]^T \in \mathbb{R}_{2 \times 1}$ is the precoding vector for the $i^{\rm th}$ stream.  A DC bias $\textbf{d}_{DC} \in \mathbb{R}_{2 \times 1}$ is added in order to ensure positive signals,  which  is required by the LEDs. Hence, the transmitted signal, $\textbf{x} \in \mathbb{R}^+_{2 \times 1}$, can be written as   
%where $\textbf{p_i}\in \textbb{R}_{2\times1}$ is the precoder vector for the $i$th stream. In addition to the three streams, a DC bias $\textbf{d}_{DC}$ is add to ensure positive LED input signals Therefore, the transmitted signal can be written as 
\begin{equation}
\mathbf{x} = \left[x_1, x_2\right]^T= \mathbf{{P}} \mathbf{s}+\mathbf{d}_{DC}
  = \sum_{i \in \{1,2,12\}}\mathbf{p}_{i}s_{i}+\mathbf{d}_{DC}
\end{equation}
and the received signal at the $k^{\rm th}$ PD, after optical-to-electrical conversion, is expressed as
\begin{equation}
y_{k}=\varsigma \zeta \mathbf{h}_{k}^{T} \mathbf{x}+{n}_{k}, \; \forall k \in \{1,2\}
\end{equation} 
where $\varsigma$ is the conversion factor of any LED, $\zeta$ is the responsivity of any PD, $\mathbf{h}_{k}=[h_{k,1}, h_{k,2}]^T$ is the DC channel gain vector between the $k^{\rm th}$ PD and the transmitting LEDs, where each element is expressed as given in (\ref{eq:fixture_app}), and $n_k \sim \mathcal{N}(0,\sigma^2_k)$ is the additive white Gaussian noise, representing the thermal and shot noise, with zero-mean and variance $\sigma^2_k$. Due to the low mobility of indoor users, we assume that the channel gains are constant during the transmission, and that perfect CSI is available at the transmitter. In order to accurately design the precoding matrix $\textbf{P}$, the following constraints need to be satisfied to ensure that the LEDs work in their dynamic range 

\begin{eqnarray}
\centering
\small
L_1(\mathbf{p}_l)&=& \sum_{i\in{1,2,12}} |p_{l,i}| \nonumber \\ &=&\text{min}\left(d_{{DC}},P_{\rm{max}}-d_{{DC}}\right), \;
\forall l \in \{1,2\}.
\end{eqnarray}

%Moreover, The constraint can be enlarged by setting $d_{DC} = P_{max}/2 \; ,\forall l$. 
%It can be noticed from (14), that the design of the precoding matrix in VLC is based on the $L_1$ norm, which is different than RF systems where the design of the precoding matrix is based on the $L_2$ norm. That is because the receiver electrical SNR in VLC is proportional to the square of the received optical power, while in \vspace{2 mm}RF systems, the received SNR is proportional to the received average power.\\   
%Users mobility in indoor VLC scenarios is very limited, therefore, we assume that CSI is fixed. 
%Signal detection at the $k$th user can be performed using  the following MMSE equalizer \cite{Ma2013}
%\begin{equation}
%g_i = \mathbf{p}^T_i \mathbf{h}_k \left(1+\mathbf{p}^T_i \,\mathbf{h}_k \, \mathbf{h}_k^T\, \mathbf{p}_k \right)^{-1},\; \forall i
%\end{equation}
%yielding the following estimated stream
%\begin{equation}
	   %   \hat{s}_i= g_i \; y_k,\; %\forall i.
%\end{equation}
The MMSE equalizer for the common stream is utilized at the $k^{\rm th}$ user for signal detection \cite{Ma2013}, followed by SIC as follows: First, user $k$ decodes the common signal $s_{12}$ while treating the other signals as noise. 
%It is to be noted that the common signal is composed of a part of the user's message in addition to that of the other user. 
Hence, the received SINR at the $k^{\rm th}$ user, for the common signal, is expressed as
\begin{equation}
\label{eq:commom}
\gamma^{12}_k = \frac{\left(\mathbf{h}^T_k \mathbf{p}_{12}\right)^2}{\left(\mathbf{h}^T_k \mathbf{p}_1\right)^2 + \left(\mathbf{h}^T_k \mathbf{p}_2\right)^2+ \hat{\sigma}^2_k}, \; \forall k \in \{1,2\}
\end{equation}
where $\hat{\sigma}^2_k=\sigma_k^2 / \left( \varsigma \zeta \right)^2$ is the normalized received noise power. For the sake of simplicity, we assume that $\varsigma \zeta=1$, and thus $\hat{\sigma}_k^2=\sigma_k^2$. 
Then, the effect of the common signal is removed using SIC. This allows for the detection of the private signal by first employing the MMSE equalizer, and then user $k$ attempts to decode its private message $s_k$, while treating the signals of other user as noise. Consequently,    
the received SINR at user $k$, for its private signal, can be written as
%After decoding the common part, SIC is performed at each user to eliminate the effect of the common stream from the received signal, and allow reliable detection for the private streams. 
%Following similar steps, each user detects its own private stream $s_k$, while considering other users’ private parts as  noise. As a result, the SINR at each user due to decoding the private stream can be written as
\begin{equation}
\label{eq:private}
\gamma^{k}_k = \frac{\left(\mathbf{h}^T_k \mathbf{p}_{k}\right)^2}{\left(\mathbf{h}^T_k \mathbf{p}_{\bar{k}}\right)^2 +  {\sigma}^2_k}, \; \forall (k,\bar{k}) \in \{(1,2),(2,1)\}
\end{equation}
and the achieved data rate at user $k$ is expressed by \cite{Mao2018}
\begin{equation}
R^{12}_k=\text{log}_{2}(1+\gamma^{12}_k),
\end{equation}
and
\begin{equation}
R_k^k=\text{log}_{2}⁡(1+\gamma^k_k),\; \forall k\in \{1,2\}    
\end{equation}
where $R_k^{12}$ and $R_k^k$ are the data rates for the common and private signals, respectively.
In order to ensure successful decoding of the common stream $s_{12}$ at both users, the common rate shall not exceed $R_{12}=\text{min}(R^{12}_1,R^{12}_2)$. The targeted common rate for each user can be achieved if $R_{12}$ is adequately shared between the two users, i.e., $R_{12}=\sum_{k=1}^2 R_{k,\rm{com}}$, where $R_{k,\rm{com}}$ is the $k^{\rm th}$ user portion of the common rate. Consequently, 
the total achievable data rate of user $k$, denoted $R_{k,\rm{ov}}$, can be expressed by \cite{Mao2018}
\begin{equation}
\label{eq:c4}
R_{k,\rm{ov}}=R_{k,\rm{com}}+R_k^k, \; \forall k \in \{1,2\}.
\end{equation}

\subsubsection{Problem Formulation}
Although conventional precoders, such as ZF and ZF-DPC, are simple and can efficiently remove MUI, they suffer from performance degradation at low SNR values. Consequently, there is a need for optimal precoding in order to maximize an objective function, e.g., sum rate, weighted sum rate (WSR), proportional fairness, or max-min fairness \cite{Pham2017}, under QoS requirements and per-LED transmit power constraints to take into account the nature of the optical signals, which are real and positive valued. Inspired by the MMSE precoding method presented in \cite{Nguyen2014}, we maximize the WSR of the studied MU-MISO VLC system. 
%This method is proven to be superior to conventional precoding techniques. 
%It would be mentioned that, the weights used in the WSR expression are selected either, based on the state of the packet queue, which leads to what is known as \textit{max-stability} service, or they may be equally chosen to maximize the sum rate, which leads to what is known as the \textit{best effort} service [80]. 
For a given weights vector $\textbf{w} =[w_1,w_2 ]$, the WSR maximization problem (P1) can be expressed as follows: 
\begin{subequations}
	\begin{align}
	\small
	\max_{\mathbf{P},\mathbf{R}_{\rm{com}}} & \quad 
	{R}(\textbf{w})={\sum_{k=1}^2}w_k \;R_{k,\rm{ov}} \tag{P1} \\
	\label{c1}
	\text{s.t.}\quad & L_1(\mathbf{p}_l) \leq \varepsilon, \; \forall l \in \{1,2\}  \nonumber \tag{P1.a}\\
	\label{c2} & \sum_{k=1}^2R_{k,\rm{com}} \leq R_{12}  \tag{P1.b}\\
	\label{c3} & \mathbf{R}_{\rm{com}}\geq \mathbf{0}  \tag{P1.d}
	\end{align}
\end{subequations}
%\vspace{-30pt}
%\begin{subequations}
%	\begin{align}
%	\small
	%\max_{\mathbf{P},\mathbf{R}_{\rm{com}}} & \quad 
	%{R}(\textbf{w})={\sum_{k=1}^2}w_k \;R_{k,\rm{ov}} \tag{P1} \\
	%\label{c1}
	%\text{s.t.}\quad & L_1(\mathbf{p}_l) \leq \varepsilon, \; \forall l \in \{1,2\}  \nonumber \tag{P1.a}\\
%	\label{c2} & \sum_{k=1}^2R_{k,\rm{com}} \leq R_{12}  \tag{P1.b}\\
%	\label{c3} & \mathbf{R}_{\rm{com}}\geq \mathbf{0}  \tag{P1.d}
%	\end{align}
%\end{subequations}
where $\textbf{R}_{\rm{com}}=[R_{1,\rm{com}},R_{2,\rm{com}}]$ is the common rate vector. (P1) is non-convex due to the presence of variables $\textbf{p}_k$ ($k \in \{1,2\}$) in the denominator of the SINR expressions (\ref{eq:commom})-(\ref{eq:private}). Thus, its solution is not straightforward. Similar to \cite{Christensen2008}, we opt for problem reformulation, where the objective becomes the minimization of the weighted MMSE, and is achieved by jointly optimizing the WMMSE precoding vectors and MSE equalizer weights. 
To obtain a local optimum, we utilize alternating optimization (AO) detailed in Algorithm \ref{Algo1} \cite{Christensen2008}, where $k$ is the iteration index, $\mathbf{w}$ is the MMSE weights vector, $\mathbf{\alpha}$ is the MSE receiver weights vector, $\mathbf{v}$ is the transformation of $\mathbf{R}_{\rm{com}} $, and $\delta$ is the tolerance threshold. In order to converge to a maximum WSR, the algorithm alternates between WMMSE precoding design and MSE equalizer weights design. For further details on the AO procedure, we refer the reader to Sections IV and V in \cite{Christensen2008}.% (\textcolor{green}{MAYBE WE NEED TO PUT MORE DETAILS ON THE SOLUTION APPROACH WITH AN ALGORITHM. YOU CAN BE INSPIRED FROM WHAT I DID IN THE OTHER PAPER.})\\
\begin{algorithm}[t]
\small{
\caption{Alternating Optimization Algorithm}
\label{Algo1}
\begin{algorithmic}[1]
\State {Initialize $k \xleftarrow{}0$, $\mathbf{P}[k]$, $R[k]$} 
\Repeat 
\State $k \xleftarrow{} k+1$; $\mathbf{P}[k-1] \xleftarrow{} \mathbf{P}$
\State Update the WMMSE weights $\mathbf{w} \xleftarrow{} \mathbf{w}(\mathbf{P}[k-1])$

\State Update the receive filter gains $\mathbf{\alpha} \xleftarrow{} \mathbf{\alpha}(\mathbf{P}[k-1])$
\State Solve (P1) using WMMSE transformation for updated ($\mathbf{w}, \mathbf{\alpha}$), then update ($\mathbf{P}$, $\mathbf{v}$)
\Until {$|{R}[k]-{R}[k-1]|\leq \delta$}.
\end{algorithmic}}
\end{algorithm}
%In each iteration, for certain users’ priority weights, the equalizer gains are firstly updated using the precoding vectors obtained from the previous stage. The second step is to update the MSE-weights. 
%Finally, by using the updated values, the new precoding vectors are obtained by solving the WMMSE problem, the process then continues until the WSR converges. According to aforementioned optimization problem, the precoder vector of the private and common streams are initialized as
%\begin{equation}
%\mathbf{P_k}= \frac{\rho P_t}{2} \ast \frac{\mathbf{h_k}}{\|\mathbf{h_k}\|}  
%\end{equation}
%and,                    %\begin{equation}
%\mathbf{P_{12}}=  (1-\rho) P_t \ast \mathbf{u_{12}}
%\end{equation}
%respectively, \\
%where, $\mathbf{u_{12}}$ is obtained from the singular value decomposition (SVD) of the channel matrix, and $\rho$ is the power allocation factor between the common and the private streams. 
Finally, the reformulated problem can be solved using  optimization software such as CVX in MATLAB\cite{CVX}. It is noted that the AO algorithm converges faster and achieves better performance than other types of precoding optimization algorithms. 
However, its complexity increases with the number of users. 
%Clearly, if more than two users exist in the system, the decoding and the encoding procedure become more complex as the optimization problem also includes determining the best decoding order for the common streams in addition to the optimum precoder. 
\subsubsection {NOMA and SDMA as Special Cases of RSMA} 
As we mentioned earlier, RSMA is a generalized MA scheme, where NOMA and SDMA are special cases. To implement SDMA from RSMA, the common stream is allocated null power, and each user's message is encoded into a private stream only. Hence, the transmitted signal in this case is
\begin{equation}
\mathbf{x}=\mathbf{P} \mathbf{s}+\mathbf{d}_{DC}=\sum_{i\in\{1,2\}}\mathbf{p}_i s_i+\mathbf{d}_{DC}
\end{equation}
and the received SINR at each user simplifies into (\ref{eq:private}).

Similarly, NOMA can be obtained from RSMA by encoding one of the users' messages as a private stream, i.e., the user with the strongest channel, and  the signal of the second user is encoded into a common stream. Assuming that user 1 has the strongest channel gain, then the transmitted signal in this case can be written as 
\begin{equation}
\mathbf{x}=\mathbf{P}\mathbf{s}+\mathbf{d}_{DC}
= \sum_{i \in \{1,12\}}\mathbf{p_i} s_i+\mathbf{d}_{DC}
\end{equation}
and the associated SINRs  of the first and second users are given by   
\begin{equation}
\gamma_1^1 = \frac{(\mathbf{h}^T_1 \mathbf{p}_{1})^2}{{\sigma}^2_1} 
\end{equation}
 and
\begin{equation}
\gamma_2^{12} = \text{min}\bigg(  \frac{(\mathbf{h}^T_1 \mathbf{p}_{12})^2}{ (\mathbf{h}^T_1 \mathbf{p}_1)^2+ {\sigma}^2_1}\;,\;\frac{(\mathbf{h}^T_2 \mathbf{p}_{12})^2}{(\mathbf{h}^T_2 \mathbf{p}_1)^2+ {\sigma}^2_2}\bigg)  \end{equation}
respectively. It is worth mentioning that the flexibility of RSMA comes at the expense of a slightly higher encoding complexity at the transmitter. 
%The increased processing complexity arises from the need to split the messages into multiple streams and then perform multiple encoding operations. 

%Table \ref{Table2} summarizes the three MA schemes and the main differences between them.

\begin{figure*}[t]

     \begin{minipage}{0.5\linewidth}
     \includegraphics[width=3in,center]{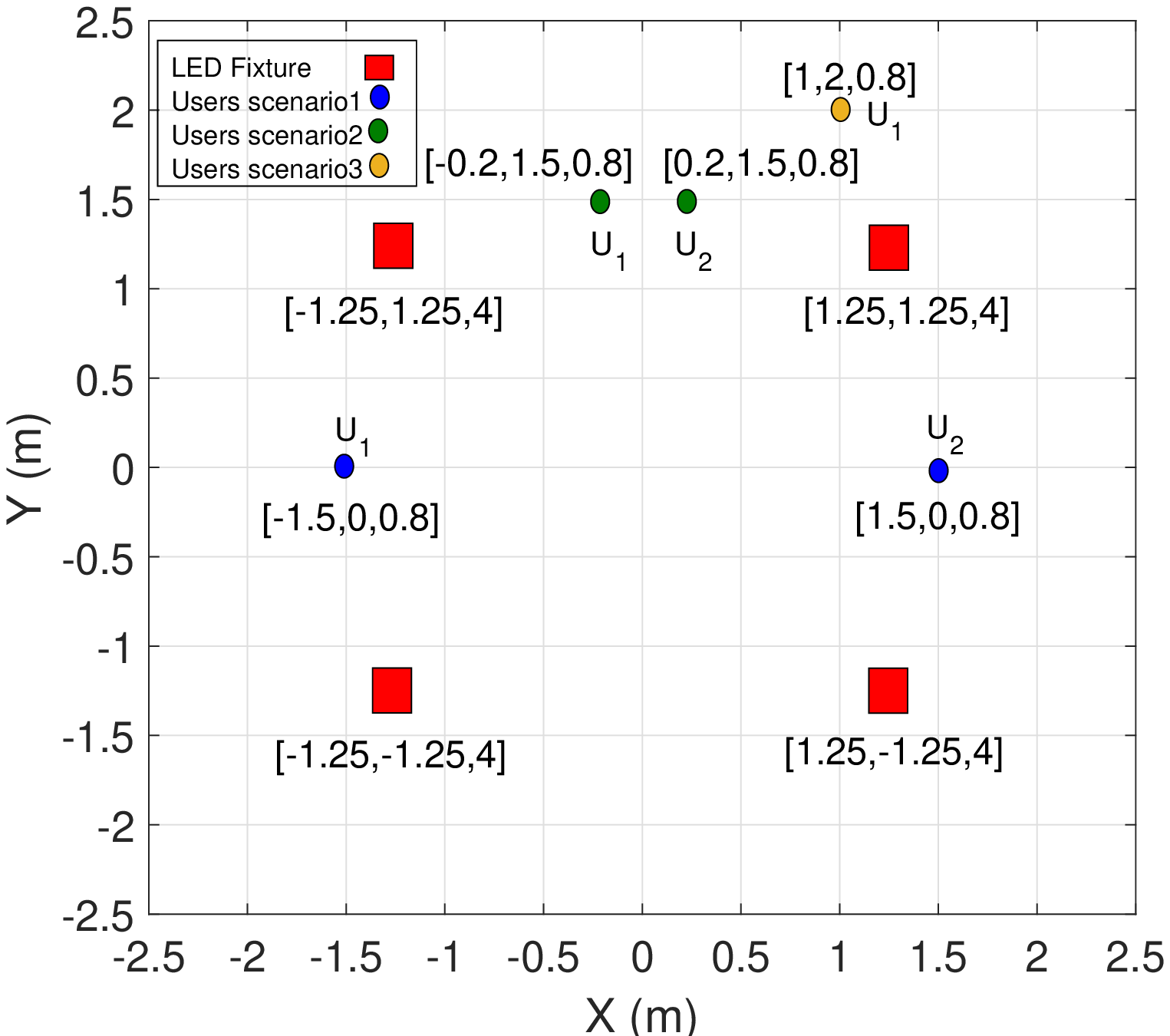}
       \caption{Room configuration and users' scenarios (4 LEDs).}
    \label{Fig:4LEDs}
     \end{minipage}
     \hfill
     \begin{minipage}{0.5\linewidth}
   \includegraphics[width=3in,center]{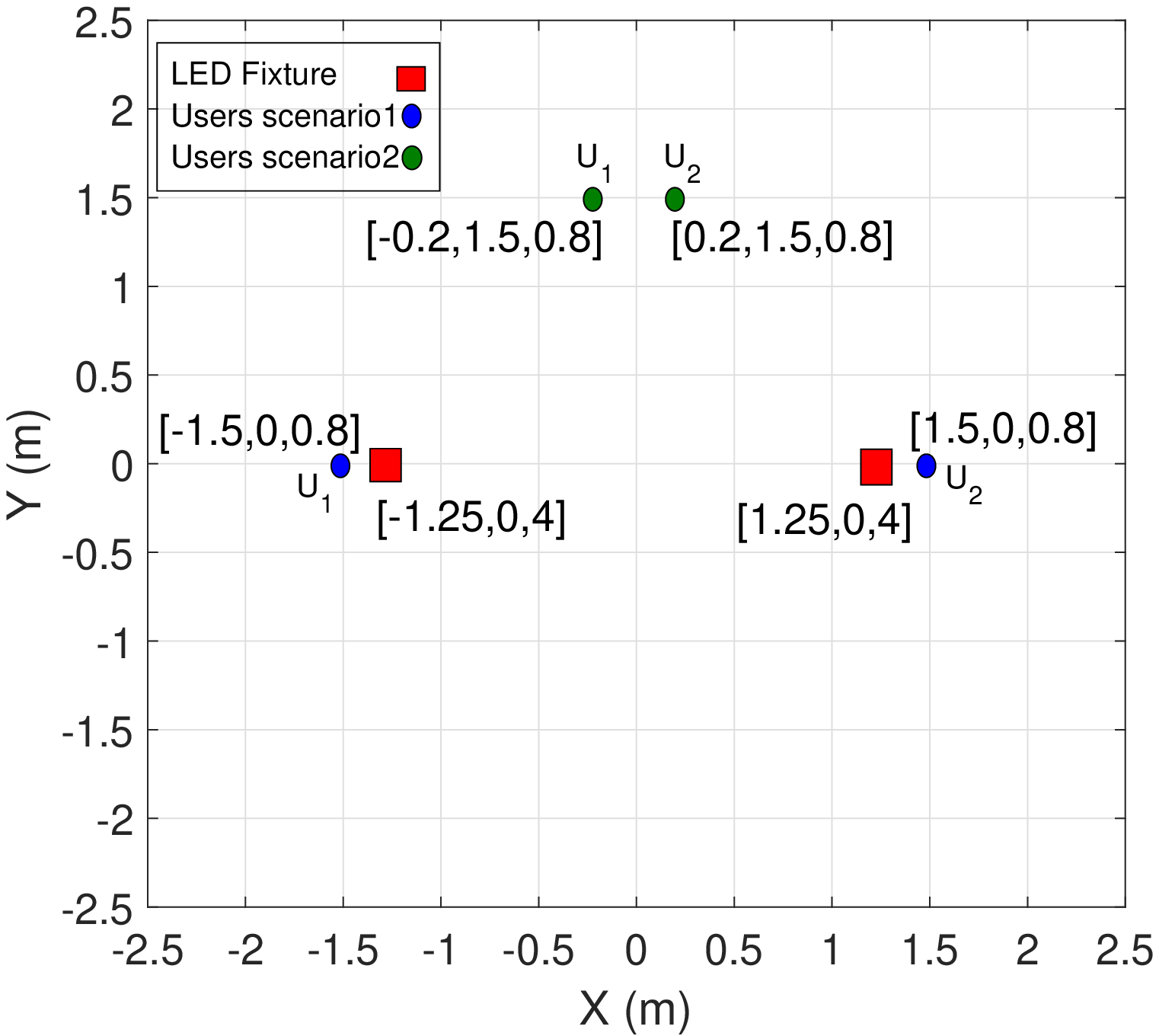}
	\caption{Room configuration and users' scenarios (2 LEDs).}
	\label{Fig:2LEDs}
     \end{minipage}
   \end{figure*}

\section {Performance Evaluation}
We present in this section different scenarios for the application of RSMA in VLC systems, where we investigate their performance in terms of WSR, and then compare them to performances of SDMA and NOMA. Moreover, we study the impact of different users' locations within an indoor space. 

%In addition to that, since VLC channels are sensitive to users’ position, we focus our study on the effect of varying users’ locations and individual channel conditions on the sum rate performance of SDMA, NOMA, and RSMA.

\begin{figure}[t]
\centering
\includegraphics[width=3in]{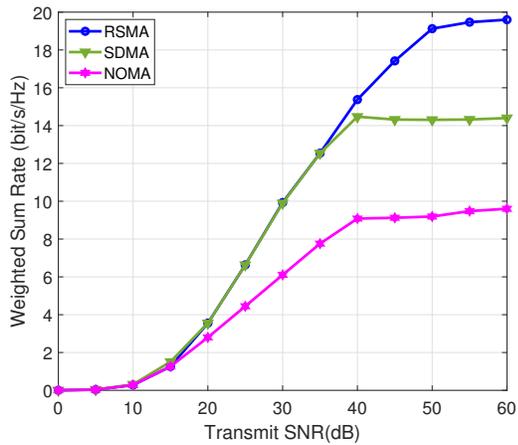}
\caption{WSR vs. SNR per LED array (Scenario 1, 4 LEDs).}
\label{Fig:Sim1}
\end{figure}

\begin{table}[t]
\scriptsize
\centering
 \caption{Simulation Parameters.}
\begin{tabular} [c]{|p{3cm}|c|c|}
\hline 
\textbf{Parameter}&\textbf{Symbol}&\textbf{Value} \\ \hline \hline
%Room size &	$5\times5\times4 \; m^3$\\ \hline
%4 LED’s array positions scenario&[ 1.25, 1.25, 4] [-1.25,-1.25,4] [-1.25,1.25,4] [1.25,-1.25,4]\\ \hline
%2 LED’s array positions scenario&	[1.25,0,4][-1.25,0,4]\\ \hline
%$1$st Users’ configuration&	[0,0,0.8],[1.5,0,0.8]\\ \hline
%$2$nd Users’ configuration&	[0,0, 0.8],[0.4,0, 0.8]\\ \hline
Number of LEDs per fixture & $Q$ &	$3600\; (60\times60)$\\ \hline
LED beam angle& $\varphi_{1/2}$ &	$60^o$\\ \hline
PD area& $A_k$ ($k=1,2$) &	$1\; cm^2$\\ \hline
Refractive index of PD& $n$ & $1.5$\\ \hline
Gain of optical filter&	$T_s(\phi_{k,i})$ ($k=1,2$) &1\\ \hline
FoV of PD & $\phi_c$	& $60^o$\\ \hline
%PD responsivity&	$0.53 A/W$\\ \hline
\end{tabular}
\label{TableIV}
\end{table}

We consider an RSMA-based MU-MISO VLC system, where two  single-PD users are served by two or four LED fixtures in a room of size $5\times5\times4 \;m^3$. 
The room configurations with the users' scenarios are detailed in Figs. \ref{Fig:4LEDs}--\ref{Fig:2LEDs}, as follows. In both figures, \textcolor{black}{main} two users' location scenarios are considered. In the first (blue circles), users are located in the middle space of the room with a separation of 3 m,  whereas in the second (green circles), users are located in the top of the room, with a smaller separation of 0.4 m. Between the two figures, the number and locations of LEDs is varied from 4 to 2. \textcolor{black}{In addition, a third scenario is considered for the 4 LEDs case, where the separation between users is 0.94 m (yellow circle for user 1 and green circle of user 2).}
All coordinates are expressed in the 3D-space system. Furthermore, we assume the same optical devices characteristics as in \cite{Ma2013}, while the two users are allocated equal priority, i.e., $w_1=w_2=\frac{1}{2}$ in the objective function of (P1). Since the noise power is assumed unity, then the SNR designates the transmit power per-LED. The remaining parameters are detailed in Table \ref{TableIV}.

%The power allocation coefficient $\rho$ was chosen to be $0.3$ in the analysis for NOMA and RSMA and clearly its value is 1 for SDMA since there is no common stream involved in it. 
%To study the effect of users' locations, we consider two users' configurations. For the first configuration, the first user position is $[0,0,0.8]$ and the second user position is $[1.5,0,0.8]$, yielding $1.5m$ separation between the two users. On the other hand, for the second configuration, the first user's coordinate is $[0, 0, 0.8]$ while the second user is located at $[0.4, 0, 0.8]$, results in $0.4m$ separation between the users. 

\begin{figure}[t]
\centering
\includegraphics[width=3in]{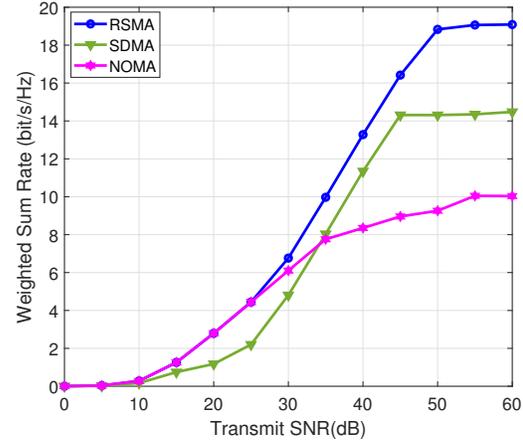}
\caption{WSR vs. SNR per LED array (Scenario 2, 4 LEDs).}
\label{Fig:Sim2}
\end{figure}

\begin{figure}[t]
\centering
\includegraphics[width=3in]{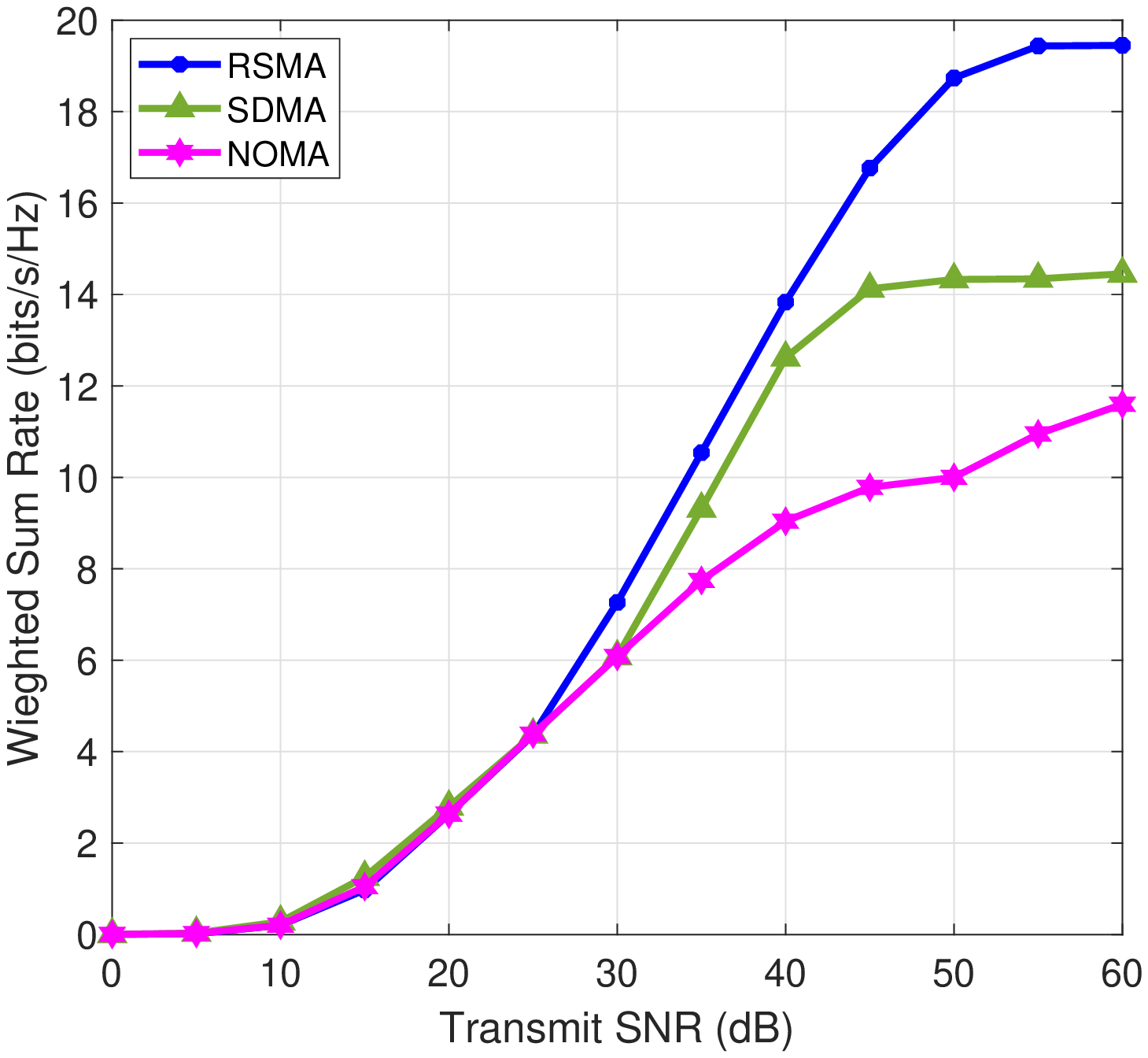}
\caption{\textcolor{black}{WSR vs. SNR per LED array (Scenario 3, 4 LEDs).}}
\label{Fig:Sim21}
\end{figure}

Fig. \ref{Fig:Sim1} shows the WSR performance for RSMA, NOMA and SDMA, in ``Scenario 1, 4 LEDs". It can be seen that RSMA outperforms both NOMA and SDMA, particularly at high SNR. In addition, SDMA performs better than NOMA, since the number of transmitter LEDs is larger than the number of users, allowing efficient management of MUI. However, SDMA performs worse than RSMA due to the difficult channels alignment between users, caused by the nature of the VLC channel. In Fig. \ref{Fig:Sim2}, the same comparison is made for ``Scenario 2, 4 LEDs". With a smaller separation between users, channels are more correlated, which is reflected by the corresponding achievable performance. For instance, using RSMA, WSR=13 bits/s/Hz (RSMA) is achieved at SNR=40 dB, compared to WSR=15.5 bits/s/Hz in Fig. \ref{Fig:Sim1}.  
Nevertheless, the performance of RSMA still exceeds that of both NOMA and SDMA. 
For SNRs below 35 dB, NOMA outperforms SDMA. Indeed, NOMA is able to distinguish different users using precoding and SIC receivers. However, for SNRs above 35 dB, this procedure is less effective, and direct beamforming using SDMA becomes privileged. Consequently, NOMA is favored at lower SNRs for low users separation, whilst SDMA is more performant for high SNRs. 
\textcolor{black}{
To examine the impact of the users separation, i.e., correlation between users channels, in Fig. \ref{Fig:Sim21},  the WSR performance has been investigated  for ``Scenario 3" (U$_1$ yellow + U$_2$ green in Fig. \ref{Fig:4LEDs}), where the users separation is 0.94 m. 
It can be seen that the WSR performance of NOMA is improved compared to ``Scenario 2." This is due to the reduced channel correlation between the two users, since the separating distance between users increased from 0.4 m to 0.94 m. Furthermore, SDMA and RSMA demonstrate improved performances, compared to ``Scenario 2," due to the lower interference between users.}

%Given that NOMA relies on the distinction between the users' channel gains, this results in a very weak performance for NOMA systems, compared to RSMA and SDMA. By noting that indoor environment implies close distance between users and given that LEDs usually located at the center of the room, this represents a major challenge for NOMA systems.

\begin{figure}[t]
\centering
\includegraphics[width=3in]{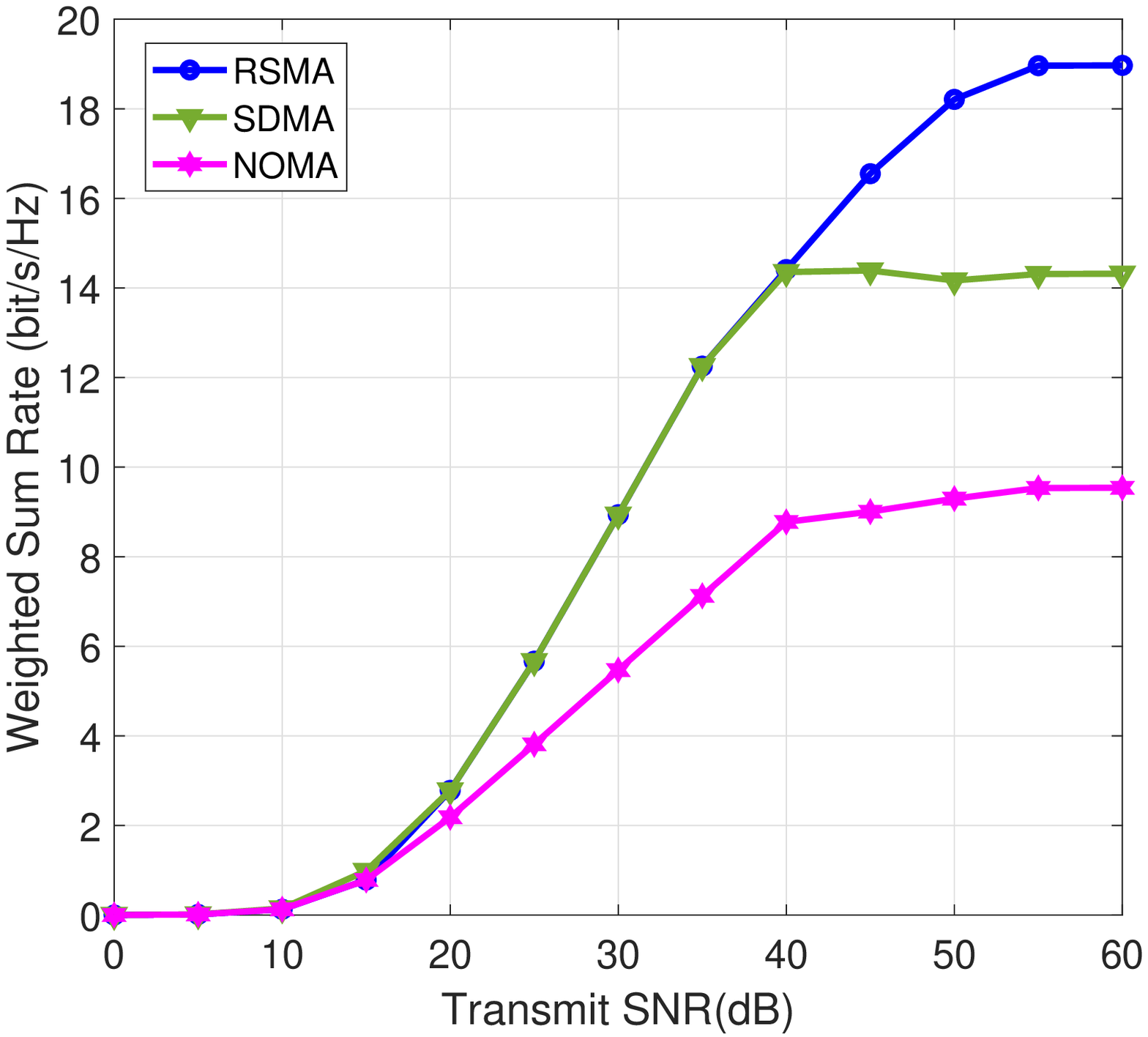}
\caption{WSR vs. SNR per LED array (Scenario 1, 2 LEDs).}
\label{Fig:Sim3}
\end{figure}

\begin{figure}[t]
\centering
\includegraphics[width=3in]{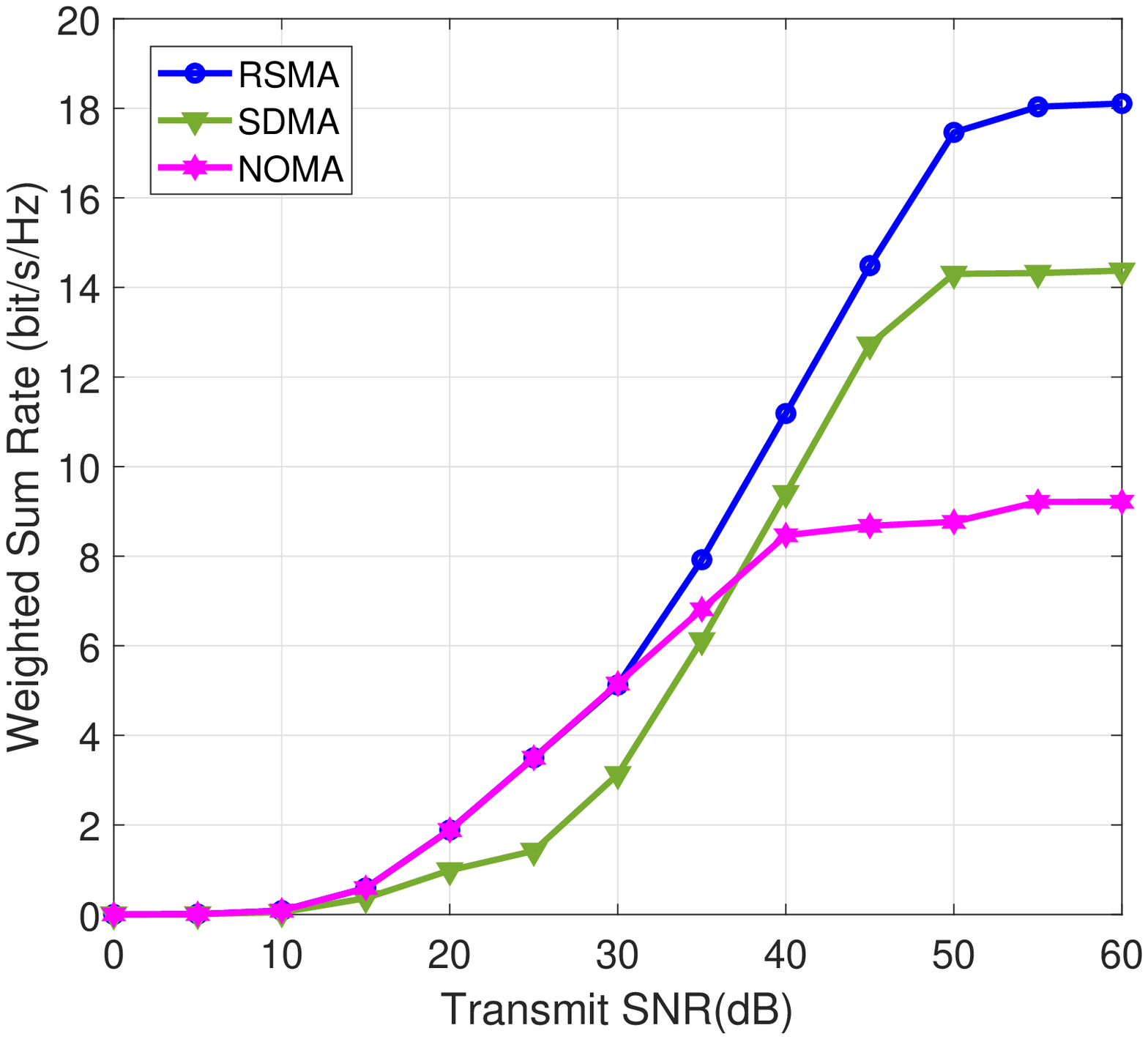}
\caption{WSR vs. SNR per LED array (Scenario 2, 2 LEDs).}
\label{Fig:Sim4}
\end{figure}

\begin{figure}[t]
\centering
\includegraphics[width=3.25in]{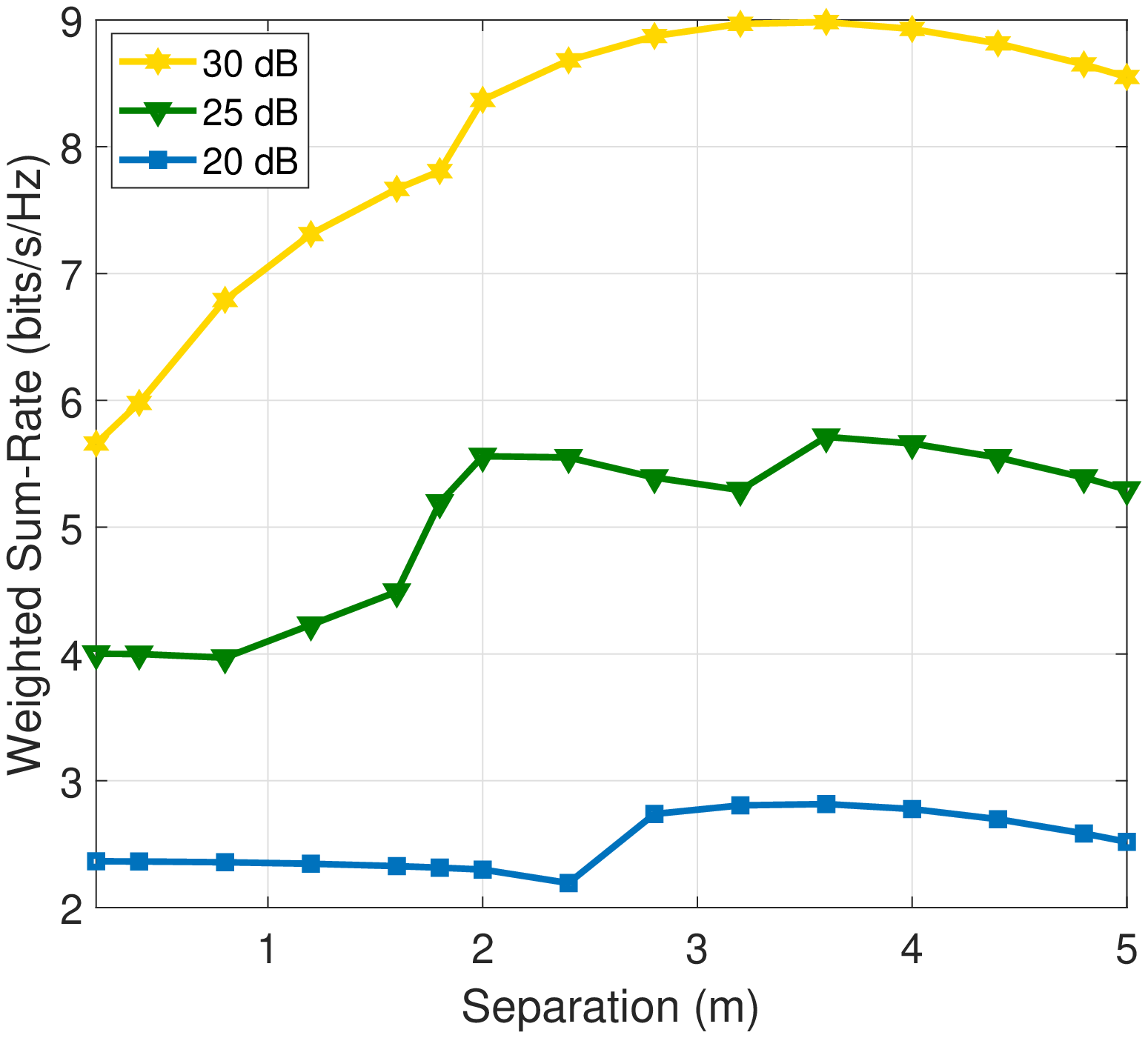}
\caption{Performance of RSMA for different users’ locations/separations and SNRs.}
\label{Fig:Sim5}
\end{figure}

In Figs. \ref{Fig:Sim3}-\ref{Fig:Sim4}, we consider the same scenarios, but with 2 LEDs. Similar to the previous results, the superiority of RSMA in terms of WSR over the other techniques is clearly illustrated. SDMA's performance is slightly degraded due to the smaller number of LEDs. Similar to Fig. \ref{Fig:Sim2}, in Fig. \ref{Fig:Sim4} the SDMA performance is degraded at SNRs below 36 dB compared to NOMA, but outperforms the latter as the SNR increases. Finally, Fig. \ref{Fig:Sim5} illustrates the users' locations impact on the WSR performance of the RSMA scheme. We consider the room setup of 2 LEDs, and two users initially located in the middle of the room. From there, the first and second users travel to the east and west walls at the same constant speed, respectively. Thus, their physical separation increases until reaching its maximum 5 m (i.e., users have reached the opposite walls). It can be seen that WSR varies with the separation, until a maximum value is achieved for a separation equal to 3.6 m. This corresponds to users locations [-1.8, 0, 0.8] and [1.8, 0, 0.8], where the correlation between channels is low, but users are very close to one of the serving LEDs to capture maximal power. However, as this separation increases above 3.6 m, the WSR degrades due to longer distances between users and LEDs. It can be seen that these optimal users' locations are the same for different SNRs. 
Consequently, designing indoor spaces using RSMA-VLC requires a careful consideration of the LEDs' and users' locations.

\section{Open Issues and Research Directions}

In this paper, we reviewed different multiple access techniques proposed to improve the spectral efficiency of VLC systems and minimize the encountered VLC-specific interference issues. Then, we addressed a preliminary work on the utilization of RSMA within VLC systems. It has been shown that RSMA is a powerful MA scheme that can provide high data rates and reliable VLC communications. However, several associated issues need to be addressed and analyzed for the practical realization of RSMA-VLC.
For instance, the impact of the non-linear distortion caused by the different circuits components, such as LEDs, PDs, and analog/digital and digital/analog converters has to be investigated. Moreover, as a novel MA technique, more efforts are required to study the design of the physical and MAC layers. Hence, different performance metrics, modulation and coding schemes, and security issues, are open research problems in the RSMA-VLC. Additionally, optimal precoding and power allocation for RSMA-VLC are still open for investigation, where new linear and non-linear techniques can be proposed and optimized. Moreover, the current literature has mainly focused on the Gaussian noise assumption, but neglected  the effect of ambient light, which can cause significant degradation in performance.

Other current assumptions include: the receiver is always pointing upward, a LoS is always available and CSI is perfectly known. However, this may not be the case in  practical scenarios, where the receiver can be differently oriented, the VLC link may experience shadowing and/or blockage, and CSI knowledge is imperfect. Consequently, the design and performance evaluation of RSMA-VLC systems that take into account these practical concerns have to be studied.
Innovative solutions to circumvent the absence of a LoS may include enabling optical cooperative communications and device-to-device (D2D) communications. Indeed, optical cooperation among VLC transmitters guarantees reliable transmissions to users in a specific area \cite{Pham2019}, whereas in D2D communications, users with strong VLC links may help forwarding data to users with blocked VLC channels \cite{Raveendran2019}. In the design of such systems, taking into consideration the different users' QoS may lead to improved performances.

%As VLC networks suffer from the problem of LOS, one could allow users to cooperate with each other in order to serve users with blocked channels which makes the application of cooperative RSMA a good solution to improve the coverage. In practical case users are usually have different QoS requirements so, it will be more realistic to consider the performance of RSMA under different users’ QoS constraints.
Finally, the analysis of massive MIMO RSMA-VLC systems is another interesting open research problem.

%Additionally, since the work assumed the use of linear precoding techniques, the use of non-linear precoding techniques and different power allocation strategies is possible. Also, a study of the optimal design for the RSMA MISO precoder based on different performance metrics such as max-min rate is still an open issue in VLC. As a trivial way to increase system capacity in VLC is to use cellular system concept hence the analysis of RSMA in cellular system is vital to improve the system performance. Another interesting line of research is the utilization and the performance of RSMA massive MIMO and cognitive radio. \\
%The work above assumed perfect channel state information at the transmitter side, a more realistic scenario is to consider the case of having channel estimation error. Moreover, most of the work in literature assumed the noise contribution only due to the thermal and shot noise, however, a more challenging job is to consider the performance of RSMA in the presence of interference from ambient light too. In addition to that, most of the analysis assumed that the receiver is pointing upward for simplicity but, a more realistic case is to consider different orientations of the receiver and its influence on the performance of RSMA and considering also the effect of shadowing. \\

\section {Conclusion}
In this paper, we provided a conceptual background of several MA schemes for MIMO-VLC systems, along with their advantages and limitations.
Specifically, our review covered NOMA and SDMA integration into VLC systems, showing how they minimize VLC interference and improve communications' performance. We reviewed also RSMA for RF systems, presented as a generalizing multiple access of NOMA and SDMA. Subsequently, we presented a preliminary study for the integration of RSMA into VLC systems, taking into consideration the per-LED power constraints. The SINR and WSR expressions are obtained for a two-user MISO VLC system, and %Channel knowledge at the transmitter was exploited in the precoder design in order to mitigate MUI and maximize the WSR performance.
results showed the flexibility of RSMA in generalizing NOMA and SDMA, at slightly increased design complexity. Through simulations, it has been proven that RSMA-VLC outperforms both techniques in terms of WSR. Also, RSMA is robust against channel correlation, and hence, it can be seen as a strong MA candidate for VLC in networks beyond 5G. Finally, a number of open issues and research directions, linked to MIMO RSMA-VLC, have been presented and discussed.

\bibliographystyle{IEEEtran}
\bibliography{IEEEabrv,tau}

\begin{IEEEbiography}
[{\includegraphics[width=1in,height=1.25in,clip,keepaspectratio]{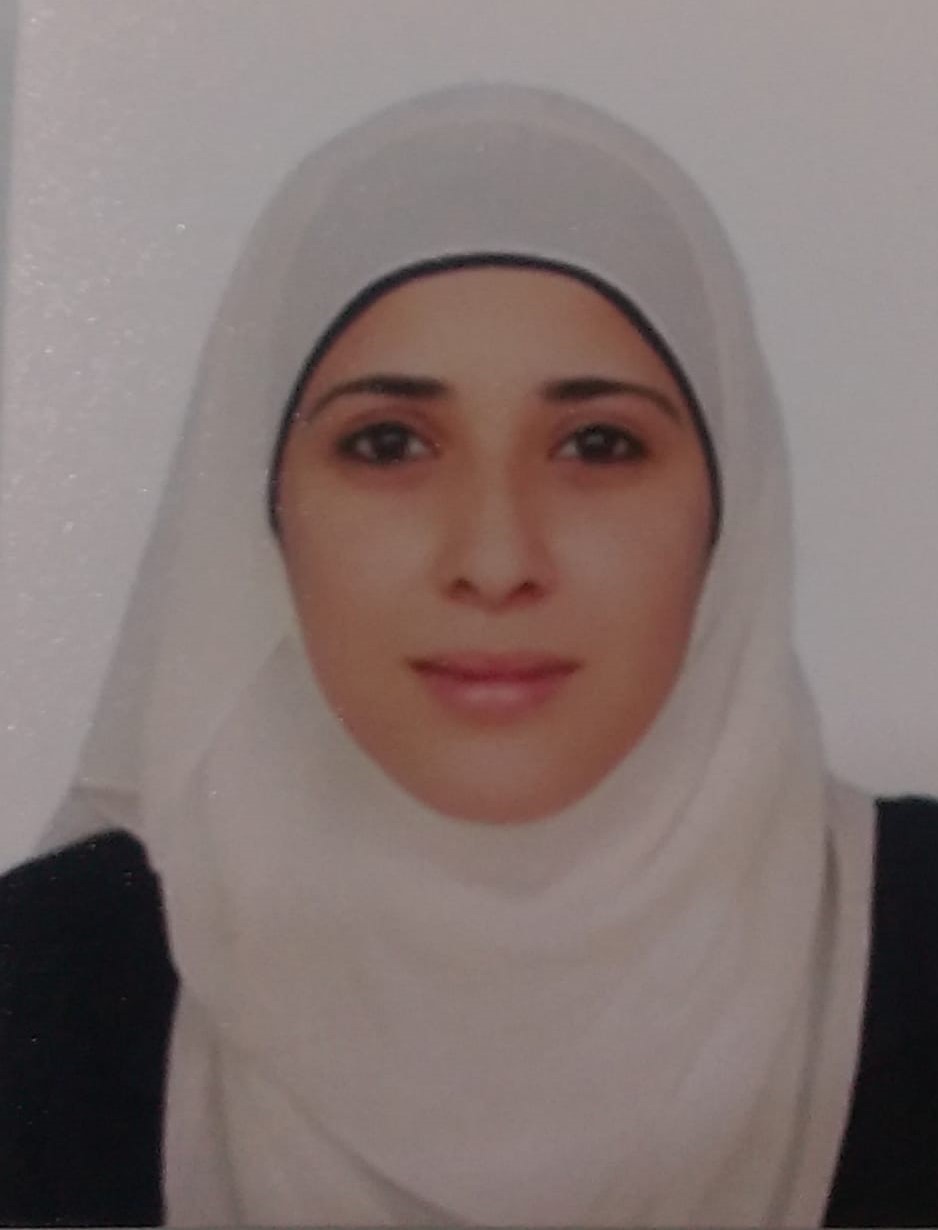}}]
{Shimaa Naser} received the M.Sc. degree in Electrical Engineering from the Jordan University of Science and Technology, Irbid, Jordan in 2015. Since 2018, she is pursuing her Ph.D. degree at the Department of Electrical and Computer Engineering of Khalifa University, Abu Dhabi, UAE. Her research interests  include  advanced  digital  signal  processing techniques for visible light communications, MIMO  techniques, orthogonal/non-orthogonal  multiple  access.
\end{IEEEbiography}

\begin{IEEEbiography}
[{\includegraphics[width=1in,height=1.25in,clip,keepaspectratio]{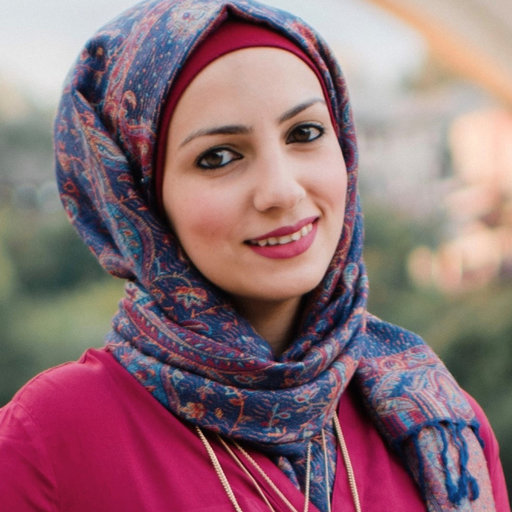}}]
{Lina Bariah} (S’13-M’19) received the M.Sc. and Ph.D degrees in communications engineering from Khalifa University, Abu Dhabi, United Arab Emirates, in 2015 and 2018. She is currently a Postdoctoral fellow with the KU Center for Cyber-Physical Systems, Khalifa University, UAE. She was a Visiting  Researcher with the Department of Systems and Computer Engineering, Carleton University, ON, Canada, in summer 2019. Her research interests include advanced digital signal processing techniques for communications, channel estimation, cooperative communications, non-orthogonal multiple access, cognitive radios, and intelligent surfaces.

\end{IEEEbiography}

\begin{IEEEbiography}
[{\includegraphics[width=1in,height=1.25in,clip,keepaspectratio]{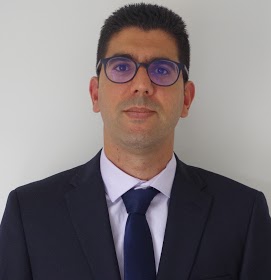}}]
{Wael Jaafar} (S’08, M’14, SM’20) received the B.Eng. degree from the Higher School of Communication (SUPCOM), Tunis, Tunisia, in 2007, and the M.A.Sc. and Ph.D. degrees in Electrical Engineering from Polytechnique Montreal, QC, Canada, in 2009 and 2014, respectively. He was a Visiting Research Intern with the Department of Computer Science, UQAM, Montreal, Canada, between February 2007 and April 2007, and a Visiting Researcher within Keio University, Tokyo, Japan, between March 2013 and June 2013, and within Khalifa University, Abu Dhabi, UAE, between November and December 2019. From 2014 to 2017, he has pursued a career in the telecommunications industry, where he has been involved in designing solutions for projects across Canada and abroad. In 2018, he held Research Fellow and Lecturer positions within the Computer Science Department of UQAM, Montreal, Canada. Since December 2018, Dr. Jaafar joined the Systems and Computer Engineering Department of Carleton University as an NSERC (National Sciences and Engineering Research Council of Canada) Postdoctoral Fellow. His current research interests include wireless communications, resource allocation, edge caching and computing and machine learning.
\end{IEEEbiography}

\begin{IEEEbiography}
[{\includegraphics[width=1in,height=1.25in,clip,keepaspectratio]{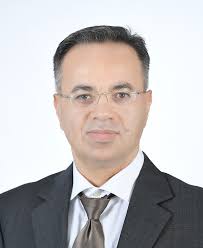}}]
{Sami Muhaidat} (S’01-M’08-SM’11) received the Ph.D. degree in electrical and computer engineering from the University of Waterloo, Waterloo, ON, Canada, in 2006. From 2007 to 2008, he was an NSERC Post-Doctoral Fellow with the Department of Electrical and Computer Engineering, University of Toronto, ON, Canada. From 2008 to 2012, he was Assistant Professor with the School of Engineering Science, Simon Fraser University, Burnaby, BC, Canada. He is
currently an Associate Professor with Khalifa University, Abu Dhabi, UAE, and a Visiting Professor with the Department of Electrical and Computer Engineering, University of Western
Ontario, London, ON, Canada. He is also a Visiting Reader with the Faculty of Engineering, University of Surrey, Guildford, U.K. Dr. Muhaidat currently serves as an Area Editor of the IEEE TRANSACTIONS ON COMMUNICATIONS, and he was previously a Senior Editor of the IEEE COMMUNICATIONS LETTERS and an Associate Editor of IEEE TRANSACTIONS ON COMMUNICATIONS, IEEE COMMUNICATIONS LETTERS and IEEE TRANSACTIONS ON VEHICULAR TECHNOLOGY. He was a recipient of several scholarships during his undergraduate and graduate studies and the winner of the 2006 NSERC PostDoctoral Fellowship Competition. Dr Muhaidat is a Senior Member IEEE.
\end{IEEEbiography}

\begin{IEEEbiography}
[{\includegraphics[width=1in,height=1.25in,clip,keepaspectratio]{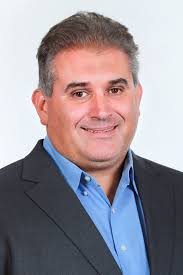}}]
{Paschalis C. Sofotasios} (S’07-M’12-SM’16) was born in Volos, Greece, in 1978. He received the M.Eng. degree from Newcastle University, U.K., in 2004, the M.Sc. degree from the University of Surrey, U.K., in 2006, and the Ph.D. degree from the University of Leeds, U.K., in 2011. He has held academic positions at the
University of Leeds, U.K., University of California at Los Angleles, CA, USA, Tampere University of Technology, Finland, Aristotle University of
Thessaloniki, Greece and Khalifa University of Science and Technology, UAE, where he currently serves as Assistant Professor in the department of Electrical Engineering and Computer Science.
His M.Sc. studies were funded by a scholarship from UK-EPSRC and his Doctoral studies were sponsored by UK-EPSRC and Pace plc. His research interests are in the broad areas of digital and optical wireless communications as well as in topics relating to special functions and
statistics. Dr. Sofotasios serves as a regular reviewer for several international journals and has been a member of the technical program
committee of numerous IEEE conferences. He currently serves as an Editor for the IEEE COMMUNICATIONS LETTERS and he received the Exemplary Reviewer Award from the IEEE COMMUNICATIONS LETTERS
in 2012 and the IEEE TRANSACTIONS ON COMMUNICATIONS in 2015 and 2016. Dr. Sofotasios is a Senior Member IEEE and he received the Best Paper Award at ICUFN 2013. 
\end{IEEEbiography}

\begin{IEEEbiography}
[{\includegraphics[width=1in,height=1.25in,clip,keepaspectratio]{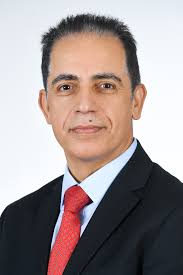}}]
{Mahmoud Al-Qutayri} (S'87, M'92, SM'02) is the Associate Dean for Graduate Studies – College of Engineering, and a Professor of Electrical and Computer Engineering at Khalifa University, UAE. He received the B.Eng., MSc and PhD degrees from Concordia University, Canada, University of Manchester, U.K., and the University of Bath, U.K., all in Electrical and Electronic Engineering in 1984, 1987, and 1992, respectively. Prior to joining Khalifa University, he was a Senior Lecturer at De Montfort University, UK. This was preceded by a Research Officer appointment at University of Bath, UK. He has published numerous technical papers in peer reviewed international journals and conferences. He coauthored a book as well as a number of book chapters. His main fields of research include embedded systems design, applications and security, design and test of mixed-signal integrated circuits, wireless sensor networks, and cognitive wireless networks. During his academic career, Dr. Al-Qutayri made many significant contributions to both undergraduate as well as graduate education. His professional service includes membership of the steering, organizing and technical program committees of many international conferences.
\end{IEEEbiography}

\begin{IEEEbiography}
[{\includegraphics[width=1in,height=1.25in,clip,keepaspectratio]{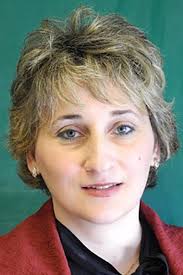}}]
{Octavia A. Dobre} (M’05–SM’07–F’20) received the Dipl. Ing. and Ph.D. degrees from Politehnica University of Bucharest (formerly Polytechnic Institute of Bucharest), Romania, in 1991 and 2000, respectively. Between 2002 and 2005, she was with New Jersey Institute of Technology, USA and Politehnica University of Bucharest. In 2005, she joined Memorial University, Canada, where she is currently a Professor and Research Chair. She was a Visiting Professor with Massachusetts Institute of Technology, USA and Universit´e de Bretagne Occidentale, France. Her research interests include enabling technologies for beyond 5G, blind signal identification and parameter estimation techniques, as
well as optical and underwater communications. She authored and co-authored over she refereed papers in these areas. Dr. Dobre serves as the Editor-in-Chief (EiC) of the IEEE Open Journal of the Communications Society and Editor of the IEEE Communications Surveys and Tutorials. She was the EiC of the IEEE Communications Letters, as well as Senior Editor, Editor, and Guest Editor for various prestigious journals and magazines. Dr. Dobre was the General Chair, Technical Program Co-Chair, Tutorial Co-Chair, and Technical CoChair of symposia at numerous conferences. shw]e was a Royal Society Scholar and a Fulbright Scholar. She obtained Best Paper Awards at various conferences, including IEEE ICC, IEEE Globecom and IEEE WCNC. Dr. Dobre is a Distinguished Lecturer of the IEEE Communications Society and
a Fellow of the Engineering Institute of Canada.
\end{IEEEbiography}

%\EOD 

\end{document}